\documentclass[aps,prd,twocolumn,superscriptaddress,showpacs,preprintnumbers,amsmath,amssymb,nofootinbib, longbibliography,
]{revtex4-2}

\usepackage{graphicx}
\graphicspath{{ps}}

\usepackage{orcidlink}

\newcommand{\res}{\ensuremath{0.22\pm0.94\pm0.42}} 
\newcommand{\xip}{\ensuremath{\xi^\prime}}
\newcommand{\xipp}{\ensuremath{\xi^{\prime\prime}}}
\newcommand{\etapp}{\ensuremath{\eta^{\prime\prime}}}
\newcommand{\alphap}{\ensuremath{\alpha^\prime/A}}
\newcommand{\betap}{\ensuremath{\beta^\prime/A}}
\newcommand{\mevcc}{\,{\ensuremath{\mathrm{\mbox{MeV}}/c^2}}}
\newcommand{\mevc}{\,{\ensuremath{\mathrm{\mbox{MeV}}/c}}}
\newcommand{\mev}{\,{\ensuremath{\mathrm{\mbox{MeV}}}}}
\newcommand{\gevcc}{\,{\ensuremath{\mathrm{\mbox{GeV}}/c^2}}}
\newcommand{\gevc}{\,{\ensuremath{\mathrm{\mbox{GeV}}/c}}}

\newcommand{\ee}{\ensuremath{e^+ e^-}}
\newcommand{\tat}{\ensuremath{\tau^+ \tau^-}}
\newcommand{\mam}{\ensuremath{\mu^+ \mu^-}}
\newcommand{\lal}{\ensuremath{\ell^+ \ell^-}}
\newcommand{\qq}{\ensuremath{q\bar{q}}}
\newcommand{\tlnn}{\ensuremath{\tau^- \to \ell^- \bar{\nu}_\ell\nu_\tau}}
\newcommand{\atlnn}{\ensuremath{\tau^+ \to \ell^+ \nu_\ell \bar{\nu}_\tau}}
\newcommand{\tmnn}{\ensuremath{\tau^- \to \mu^- \bar{\nu}_\mu\nu_\tau}}
\newcommand{\tmnng}{\ensuremath{\tau^- \to \mu^- \bar{\nu}_\mu\nu_\tau\gamma}}
\newcommand{\menn}{\ensuremath{\mu^- \to e^- \bar{\nu}_e\nu_\mu}}
\newcommand{\kpipi}{\ensuremath{K^- \to \pi^-\pi^0}}
\newcommand{\kpipipi}{\ensuremath{K^- \to \pi^-\pi^0\pi^0}}
\newcommand{\kcpi}{\ensuremath{K^- \to \pi^-\pi^+\pi^-}}
\newcommand{\kmunu}{\ensuremath{K^- \to \mu^- \bar{\nu}_{\mu}}}
\newcommand{\kpmunu}{\ensuremath{K^- \to \pi^0 \mu^- \bar{\nu}_{\mu}}}
\newcommand{\kpenu}{\ensuremath{K^- \to \pi^0 e^- \bar{\nu}_{e}}}
\newcommand{\pimunu}{\ensuremath{\pi^- \to \mu^- \bar{\nu}_{\mu}}}
\newcommand{\cascade}{\ensuremath{\tau^- \to \mu^- (\to e^- \bar{\nu}_e\nu_\mu) \bar{\nu}_\mu\nu_\tau}}
\newcommand{\mennf}{\ensuremath{\mu\to e\nu\nu}}
\newcommand{\tmnnf}{\ensuremath{\tau\to \mu\nu\nu}}

\begin{document}

\title{\quad \\[0.5cm] {Study of the muon decay-in-flight in the {\boldmath{$\tmnn$}} decay to measure the Michel parameter {\boldmath{\xip}}}}

\noaffiliation
  \author{D.~Bodrov\,\orcidlink{0000-0001-5279-4787}} 
  \author{P.~Pakhlov\,\orcidlink{0000-0001-7426-4824}} 
  \author{I.~Adachi\,\orcidlink{0000-0003-2287-0173}} 
  \author{H.~Aihara\,\orcidlink{0000-0002-1907-5964}} 
  \author{S.~Al~Said\,\orcidlink{0000-0002-4895-3869}} 
  \author{D.~M.~Asner\,\orcidlink{0000-0002-1586-5790}} 
  \author{H.~Atmacan\,\orcidlink{0000-0003-2435-501X}} 
  \author{T.~Aushev\,\orcidlink{0000-0002-6347-7055}} 
  \author{R.~Ayad\,\orcidlink{0000-0003-3466-9290}} 
  \author{V.~Babu\,\orcidlink{0000-0003-0419-6912}} 
% ...
\collaboration{The Belle Collaboration}

\begin{abstract}
We present the first measurement of the Michel parameter $\xip$ in the $\tmnn$ decay using the full data sample of $988\,\text{fb}^{-1}$ collected by the Belle detector operating at the KEKB asymmetric energy $\ee$ collider.
The method is based on the reconstruction of the $\menn$ decay-in-flight in the Belle central drift chamber and relies on the correlation between muon spin and its daughter electron momentum. We study the main sources of the background that can imitate the signal decay, such as kaon and pion decays-in-flight and charged particle scattering on the detector material. Highly efficient methods of their suppression are developed and applied to select 165 signal-candidate events. We obtain $\xip=\res$ where the first uncertainty is statistical, and the second one is systematic. The result is in agreement with the Standard Model prediction of $\xip=1$.
\end{abstract}

\maketitle
\tighten

\section{Introduction}
In the Standard Model (SM), the $\tau$ lepton decay proceeds through a weak charged current, whose amplitude can be approximated with high accuracy by the four-fermion interaction with the $V-A$ Lorentz structure. A deviation from this structure would indicate physics beyond the SM, which can be caused by an anomalous coupling of the $W$ boson with the $\tau$ lepton, a new gauge or charged Higgs bosons contribution, etc~\cite{Bryman:2021teu,Herczeg:1985cx,Krawczyk:2004na,Chun:2016hzs}. The presence of massive neutrinos can also modify experimental observables, leading to a deviation from the SM prediction~\cite{Marquez:2022bpg}.

The most general form of the Lorentz invariant, local, derivative-free, lepton-number-conserving four-fermion interaction Hamiltonian~\cite{Michel:1949qe} leads to the following matrix element of the $\tlnn$\footnote{Charge conjugation is implied throughout the paper unless otherwise indicated.} decay ($\ell=e$ or $\mu$) written in the form of helicity projections~\cite{Scheck:1984md,Mursula:1984zb,Fetscher:1986uj}:
\begin{eqnarray} \label{eq:1.1}
M=\dfrac{4G_F}{\sqrt{2}}\sum_{\substack{\lambda=S,V,T\\
\varepsilon,\omega=L,R} }g^\lambda_{\varepsilon\omega}\left\langle \bar {\ell}_\varepsilon \left| \Gamma^\lambda \right| (\nu_\ell)_\alpha \right\rangle
\left\langle (\bar \nu_\tau)_\beta \left|
\Gamma_\lambda \right| \tau_\omega \right\rangle , 
\end{eqnarray}
where
\begin{eqnarray} \label{eq:1.2}
\Gamma^S=1, \quad\Gamma^V=\gamma^\mu, \quad \Gamma^T=\dfrac{i}{2\sqrt{2}}(\gamma^\mu\gamma^\nu-\gamma^\nu\gamma^\mu);
\end{eqnarray}
$S$, $V$, and $T$ denote scalar, vector, and tensor interaction, respectively; $\varepsilon,\omega=L,R$ means left- and right-handed leptons, respectively. Each set of indices $\lambda$, $\varepsilon$, and $\omega$ uniquely determines the neutrino handedness $\alpha$ and $\beta$. The total strength of the weak interaction in Eq.~\eqref{eq:1.1} is given by $G_F$, while $g^\lambda_{\varepsilon\omega}$ are normalized as
\begin{eqnarray}\label{eq:1.3}
\sum_{\substack{\varepsilon,\omega=L,R} }\left(\frac{1}{4}|g^S_{\varepsilon\omega}|^2+|g^V_{\varepsilon\omega}|^2+3|g^T_{\varepsilon\omega}|^2\right)\equiv 1.
\end{eqnarray}

It is convenient to express the observables in the lepton decay in terms of the Michel parameters (MPs), which are bilinear combinations of the coupling constants $g^\lambda_{\varepsilon\omega}$. The MPs are described in detail elsewhere~\cite{Fetscher:1993sxo}.

At present, in $\tau$ decays four Michel parameters, $\rho$, $\eta$, $\xi$, and $\xi\delta$, have been measured with accuracies at the level of a few percent~\cite{Workman:2022ynf}, and the obtained values $\rho=0.745\pm0.008$, $\eta=0.013\pm0.020$, $\xi=0.985\pm0.030$, and $\xi\delta=0.746\pm0.021$ are in agreement with the SM prediction of $\rho=3/4$, $\eta=0$, $\xi=1$, and $\xi\delta=3/4$. These parameters describe the differential decay width, integrated over the neutrinos momenta and summed over the daughter lepton spin. Measurements of the remaining Michel parameters, \xip, \xipp, \etapp, \alphap, and \betap, requires knowledge of the daughter lepton polarization, and no measurements of them have yet been performed. The only exception is two parameters, $\xi\kappa$ and $\bar\eta$, obtained in the radiative leptonic $\tau$ decays by the Belle collaboration~\cite{Shimizu:2017dpq}. These parameters are related to the Michel parameters $\xip$ and $\xipp$ through linear combinations with the parameters $\xi$, $\xi\delta$, and $\rho$: $\xip=-\xi-4\xi\kappa+8/ 3 \xi \delta$ and $\xipp=16/3 \rho- 4\bar{\eta}-3$. Substituting parameters $\xi$ and $\xi\delta$ with their SM values and $\xi\kappa$ with the value for the radiative muonic $\tau$ decay from Ref.~\cite{Shimizu:2017dpq}, one obtains $\xip=-2.2\pm2.4$. However, this measurement still suffers from very large uncertainties: physically allowed $\xip$ values range from $-1$ to $1$ (in SM, it is equal to $1$).

In this paper, we present the first direct measurement of the Michel parameter $\xip$ in the $\tmnn$ decay. This parameter determines the longitudinal polarization of muons $P_L$ and enters the term of the $\tmnn$ differential decay width that does not depend on the $\tau$ lepton polarization. The parameter $\xip$ is written in terms of the coupling constants $g^\lambda_{\varepsilon\omega}$ as
\begin{eqnarray}\label{eq:1.4}
\xip=1-2\sum_{\substack{\omega=L,R} }\left(\frac{1}{4}|g^S_{R\omega}|^2+|g^V_{R\omega}|^2+3|g^T_{R\omega}|^2\right).
\end{eqnarray}
Thus, a measurement of $\xip$ provides the necessary information
required to calculate the probability of an unpolarized $\tau$ lepton to
decay to a right-handed muon: $Q^\mu_{R}=(1-\xip)/2$. This paper is accompanied by a Letter in Physical Review Letters~\cite{Belle:2023udc}.

\section{Method}
\subsection{Differential decay width}

The method of the muon polarization measurement is based on the $\menn$ decay reconstruction since the electron momentum in the muon rest frame correlates with the muon spin. Initially, the idea was suggested in Ref.~\cite{Fetscher:1990su}, where it was proposed to use stopped muons. Recently, it was proposed to use the muon decay-in-flight (kink) in the tracking system of the detector to measure $\xip$ in the $\tmnn$ decay in a future experiment at the Super Charm-Tau Factory~\cite{Bodrov:2021vkn, Bodrov:2021hfe}. In this paper, we rely on the adaptation of this method for the application at the $B$-factories from Ref.~\cite{Bodrov:2022mbd}. 

The differential decay width of the cascade decay $\cascade$ obtained in Ref.~\cite{Bodrov:2022mbd} follows
\begin{eqnarray}\label{eq:2.1}
\dfrac{d^3\Gamma}{d x\,d y \,d\!\cos{\theta_e}} = \mathcal{B}_{\mennf}\dfrac{12\Gamma_{\tmnnf}}{1-3x_0^2} y^2\sqrt{x^2-x_0^2}\nonumber\\
\times\left[(3-2y)F_{IS}(x)+ (2y-1)F_{IP}(x)
\cos{\theta_e}\right].
\end{eqnarray}
Here $\Gamma_{\tmnnf}$ is the partial width of the $\tmnn$ decay; $\mathcal{B}_{\mennf}$ is the branching fraction of the $\menn$ decay; $x=E_\mu/W_{\mu\tau}$ is the reduced muon energy in the $\tau$ rest frame [$W_{\mu\tau}=(m^2_\mu + m^2_\tau)/(2m_\tau)$ is the maximum muon energy]; $x_0 = m_\mu/W_{\mu\tau}$ is the reduced muon mass; and $y=2E_e/m_\mu$ is the ratio of the electron energy to its maximum value in the muon rest frame. Functions $F_{IS}(x)$ and $F_{IP}(x)$ are expressed in terms of Michel parameters and depend only on $x$:
\begin{eqnarray}\label{eq:2.2}
\begin{aligned}
F_{IS}(x)  &= x(1-x)+\dfrac{2}{9}\rho(4x^2-3x-x_0^2)+\eta x_0(1-x),\\
F_{IP}(x) &=  \dfrac{1}{54}\sqrt{x^2-x_0^2}\left[-9\xip\left(2x-3+\dfrac{x_0^2}{2}\right)\right.\\
&\qquad\qquad\quad\left.+4\xi\left(\delta-\dfrac{3}{4}\right)\left(4x-3-\dfrac{x_0^2}{2}\right)\right].
\end{aligned}
\end{eqnarray}
Since $\rho$, $\eta$, $\xi$, and $\xi\delta$ are measured very precisely, we fix their values to the SM expectations;\footnote{It is checked that this assumption has a negligible effect.} thus, only \xip\ in Eqs.~\eqref{eq:2.1} and \eqref{eq:2.2} is to be determined.

The variable $\theta_e$ is the angle between $\vec{n}_{\mu}$ and $\vec{n}_e'$, where $\vec{n}_{\mu}$ is the direction opposite to the $\tau$ lepton momentum in the muon rest frame at the muon production vertex, and $\vec{n}_e'$ is the direction of the electron in the muon rest frame at the muon decay vertex. The former vector is represented in the conventional coordinate system introduced in Ref.~\cite{Bodrov:2022mbd} as $(\bar{x}_1,~\bar{x}_2,~\bar{x}_3)$ while $\vec{n}_e'$ is represented in the coordinate system obtained from the initial one by rotation through an angle $\phi$ (the muon momentum angle of rotation in the magnetic field of the Belle detector before the decay). The procedure of the coordinate system rotation and $\theta_e$ calculation is explained in detail in Ref.~\cite{Bodrov:2022mbd}. 

The angle $\theta_e$ has a simple physical meaning when the muon decays immediately at the production vertex: this is the angle between mother and daughter charged leptons in the muon rest frame. Once the muon propagates in the magnetic field of the detector, its momentum in the laboratory frame and spin in the muon rest frame are rotated through the same angle $\phi$ (assuming $g_\mu-2\approx 0$ without loss of precision~\cite{Workman:2022ynf}). The rotation of the coordinate system in each event is designed to compensate for the effect of the magnetic field, bringing the event to the case of the instantaneous muon decay.

\subsection{\boldmath{$\tau$} lepton momentum reconstruction}

For the $\xip$ measurement, a knowledge of the $\tau$ lepton momentum is essential. While the $\tau$ energy is known from the beam energy (up to the initial-state radiation), it is not feasible to reconstruct the true direction of the $\tau$ momentum due to neutrinos in the final state. However, it is possible to find the region where the $\tau$ lepton momentum is directed using the second (tagging) $\tau$ lepton in the event~\cite{Kuhn:1993ra}. The method is based on the kinematics of the $\tat$-pair production and decay in the center-of-mass (c.m.) frame. 

For hadronic modes of the tagging $\tau$, the angle between $\tau$ lepton and daughter hadron momenta is the following:
\begin{eqnarray}\label{eq:2.3}
    \cos{\psi}=\dfrac{2E_\tau E_h-m_\tau^2-m_h^2}{2p_\tau p_h}.
\end{eqnarray}
Here $E_\tau$ and $p_\tau$ are the $\tau$ lepton energy and momentum magnitude in the c.m. frame; $E_h$, $p_h$, and $m_h$ are the hadron system energy, momentum magnitude, and invariant mass, respectively. We use all one-prong and three-prong modes of the tagging $\tau$, including $\atlnn$ ($\ell=e$ or $\mu$); however, we treat the leptonic mode as a hadronic one for simplification. 

For the signal $\tmnn$ decay, the angle between $\tau$ lepton and daughter muon momenta is restricted to
\begin{eqnarray}\label{eq:2.4}
    \dfrac{2E_\tau E_\mu-m_\tau^2-m_\mu^2}{2p_\tau p_\mu}\leq\cos{\chi}\leq\dfrac{E_\tau E_\mu-m_\tau m_\mu}{p_\tau p_\mu}.
\end{eqnarray}
Here, $E_\mu$ and $p_\mu$ are the muon energy and momentum in the c.m. frame.

\begin{figure}[htb]
    \centering
    \includegraphics[width=1\linewidth]{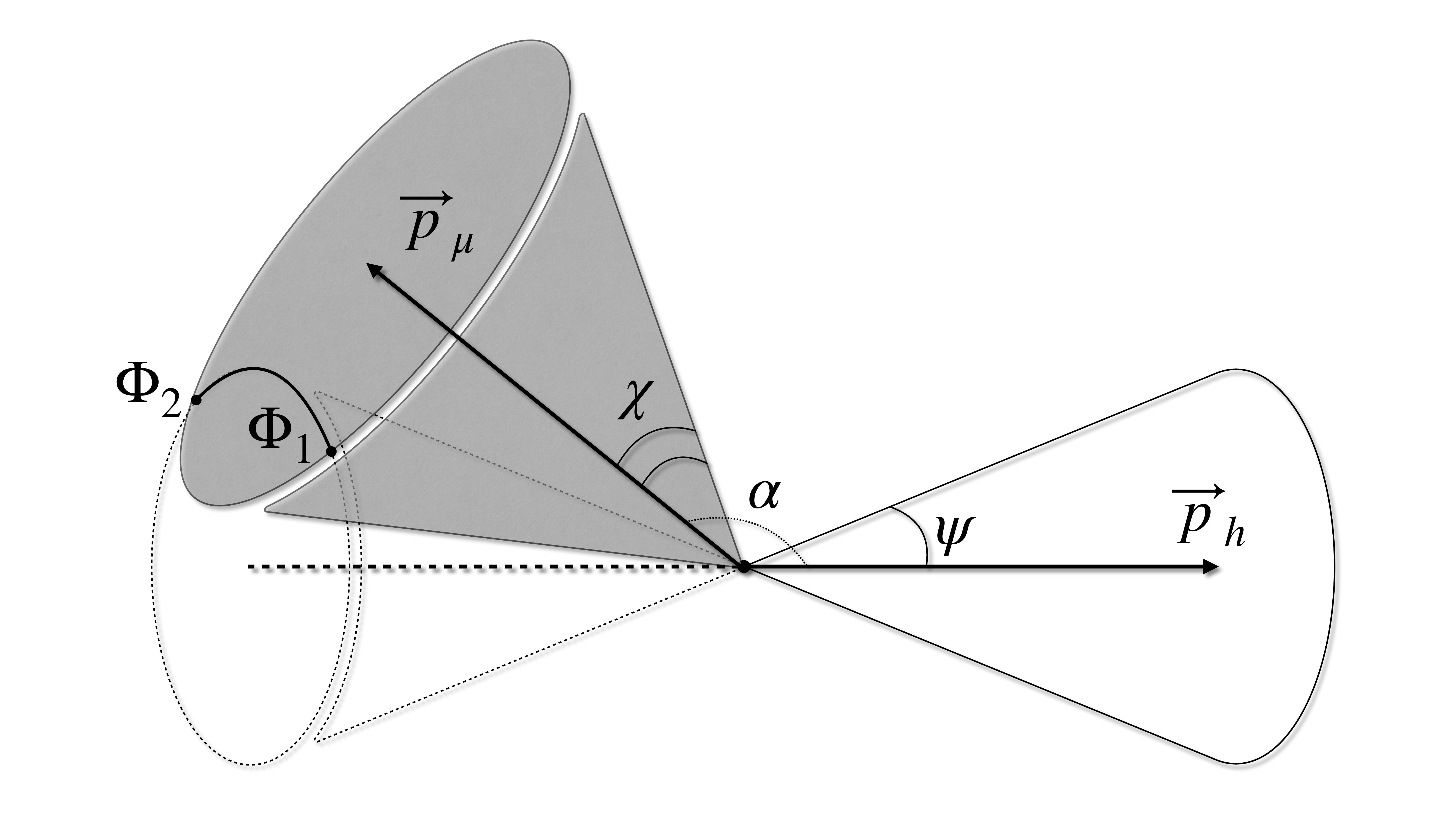}
    \caption{Geometric interpretation of the $\tau$ lepton momentum reconstruction.}\label{fig:2.1}
\end{figure}

Thus, the true $\tat$ pair production axis lies on the generatrix of a cone with an apex angle of $2\psi$ and inside a cone with an apex angle of $2\chi$ (see Fig.~\ref{fig:2.1}). 
This restricts the region of possible $\tau$ lepton directions to an arc $(\Phi_1,\,\Phi_2)$, where $\Phi_1$ and $\Phi_2$ are defined as follows
\begin{eqnarray}\label{eq:2.5}
\begin{aligned}
   \Phi_1&=\pi+\arcsin{\dfrac{\cos{\psi}\cos{\alpha}+\cos{\chi}}{\sin{\psi}\sin{\alpha}}},\\
   \Phi_2&=2\pi-\arcsin{\dfrac{\cos{\psi}\cos{\alpha}+\cos{\chi}}{\sin{\psi}\sin{\alpha}}}.
\end{aligned}
\end{eqnarray}
Here $\alpha$ is an angle between $\vec{p}_\mu$ and $\vec{p}_h$. For simplicity, we use the average value of $\Phi=(\Phi_1 + \Phi_2)/2=-\pi/2$ instead of averaging Eq.~\eqref{eq:2.1} over $(\Phi_1,\,\Phi_2)$. This approximation has a negligible impact on the $\xip$ measurement: the increase of the statistical uncertainty is less than $0.01$.

\subsection{Decay vertex reconstruction}

Track reconstruction at Belle is optimized for long-lived particles that originate close to the interaction point of the beams (IP) and does not contain dedicated algorithms for identifying charged particle decays in flight. However, tracks that do not point to IP are also reconstructed with a considerable efficiency. In our case, the muon track is reconstructed first, and it may absorb some hits produced by the daughter electron, thereby smearing the muon momentum resolution. The remaining hits are used to reconstruct the electron track.

We define the decay vertex as a point of the closest approach of muon and electron tracks in the region of their endpoint and starting point, respectively. 

\section{The data sample and the Belle detector}

This analysis is based on a data sample taken at or near the $\Upsilon(1S)$, $\Upsilon(2S)$, $\Upsilon(3S)$, $\Upsilon(4S)$, and $\Upsilon(5S)$ resonances with an integrated luminosity of $988\,\text{fb}^{-1}$ corresponding to about $912 \times 10^6$ \tat\ pairs~\cite{Belle:2012iwr}. The data are collected with the Belle detector~\cite{Belle:2000cnh} at the KEKB asymmetric-energy \ee\ collider~\cite{Kurokawa:2001nw, *[][{ and references therein.}] Abe:2013kxa}.

The Belle detector is a large-solid-angle magnetic
spectrometer that consists of a silicon vertex detector (SVD), a 50-layer central drift chamber (CDC), an array of
aerogel threshold Cherenkov counters (ACC), a barrel-like arrangement of time-of-flight scintillation counters (TOF), and an electromagnetic calorimeter comprised of CsI(Tl) crystals (ECL) located inside a super-conducting solenoid coil that provides a 1.5~T magnetic field. An iron flux-return located outside of the coil is instrumented to detect $K_L^0$ mesons and to identify muons (KLM).

The most critical subdetector for this study is CDC~\cite{Hirano:2000ei}. It has the following dimensions: the length is $2400\,\text{mm}$, and the inner and outer radii are $83$ and $874\,\text{mm}$, respectively. This size is large enough to reliably reconstruct both daughter electron and mother muon tracks.

To study the background processes, optimize the selection criteria, and obtain the fit function, signal and background Monte Carlo (MC) samples are used. 

A signal MC sample of $\ee\to\tat$ with the following $\cascade$ cascade decay is $\sim50$ times larger than the data. The production and subsequent decay of $\tat$ pairs are generated with {\footnotesize KKMC}~\cite{Jadach:1999vf} and {\footnotesize TAUOLA}~\cite{Was:2000st, Jadach:1993hs} generators, respectively, and decay products are propagated by {\footnotesize GEANT3}~\cite{Brun:1987ma} to simulate the detector response. The $\menn$ decay is also generated by {\footnotesize GEANT3}, assuming muons are unpolarized as if $\xip=0$. To speed up the signal MC sample generation, we reduce the muon lifetime in {\footnotesize GEANT3} by 100. This procedure is justified and only slightly biases the distribution of the muon decay length because the CDC size is much smaller than the average flight distance of the muon from the $\tau$ decay, which is of the order of a kilometer. We evaluate the effect of this reduction of the muon lifetime in the MC generation and quote a systematic uncertainty associated with it.

\begin{figure}[htb]
  \centering
  \includegraphics[width=1\linewidth]{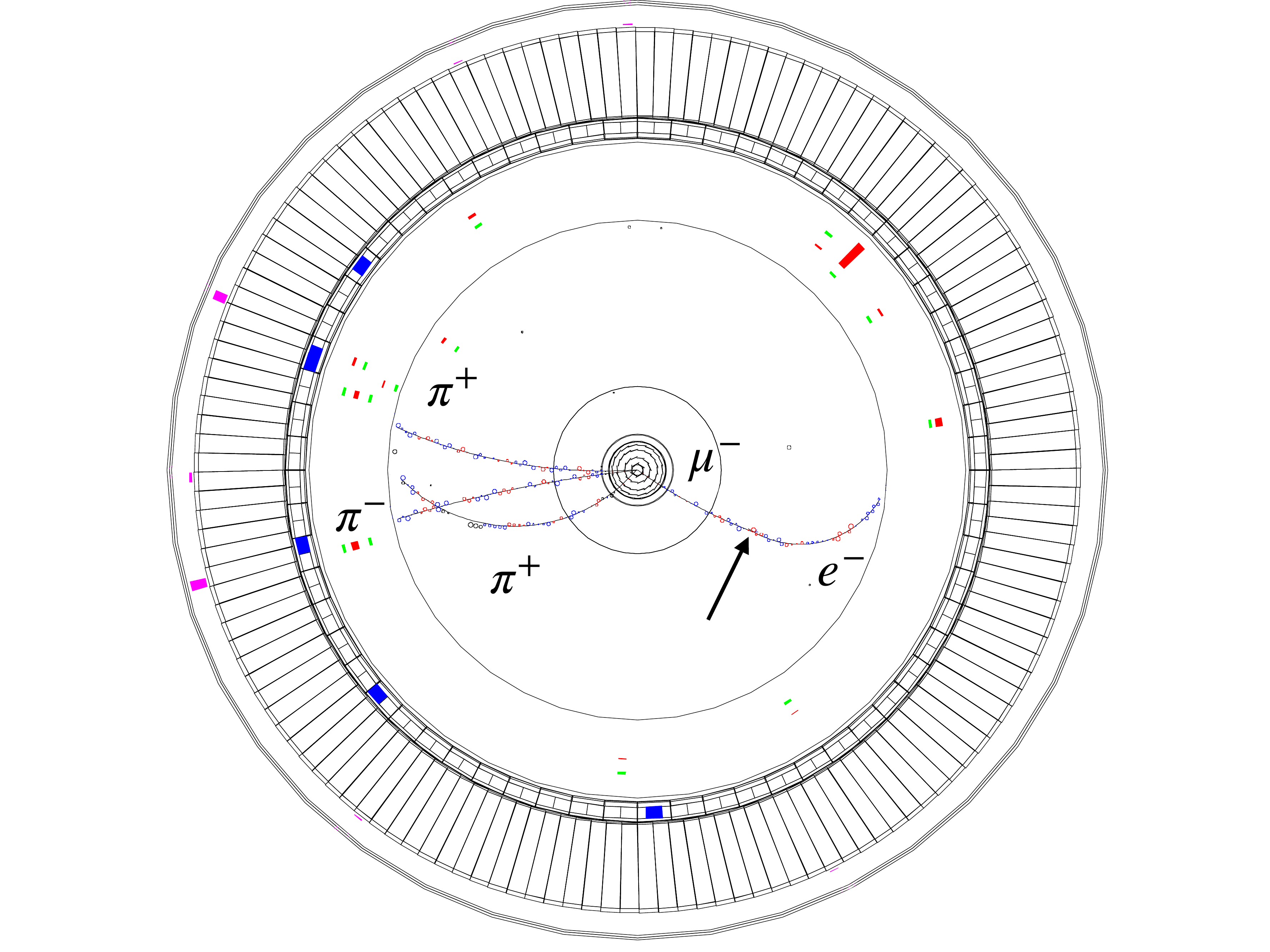}
\caption{Event display of a MC event $\ee\to\tat\to(\pi^+\pi^-\pi^+\bar{\nu}_\tau)(\mu^-\bar\nu_\mu\nu_\tau)$ with $\menn$ decay in the CDC (the arrow points to the decay vertex). The Belle detector, without the KLM, is shown projected onto $x$--$y$ plane.}
\label{fig:3.1} 
\end{figure}

An example of the $\ee\to\tat$ MC event display with the $\menn$ kink in the CDC is presented in Fig.~\ref{fig:3.1}.
The $\menn$ decay that occurred in the central volume of the CDC is clearly observed as a kinked track due to the change of the trajectory curvature since the daughter electron from the $\menn$ decay has a smaller momentum in the laboratory frame compared to the mother muon. Both electron and muon trajectories are reconstructed as separate tracks by the Belle track reconstruction algorithm.

The background consists of $\tat$-pair events
without a $\menn$ decay and non-$\tat$-pair events. The MC sample for the former contribution is generated the same way as the signal, with an exception of the $\menn$ decay generation step. The non-$\tat$-pair background consists of the dimuon $\ee\to\mam$ process, $\ee\to \qq$ ($q=u,\,d,\,s$, and $c$) continuum and $\ee\to\Upsilon(4S)\to B\bar{B}$ events, two-photon mediated processes ($\ee\to\ee\lal,\,\ee\qq$, where $\ell=e,\,\mu$ and $q=u,\,d,\,s$, and $c$), and Bhabha scattering generated with {\footnotesize KKMC}, {\footnotesize EvtGen}~\cite{Lange:2001uf}, {\footnotesize AAFH}~\cite{Berends:1986ig}, and {\footnotesize BHLUMI}~\cite{Jadach:1991by} generators, respectively.
Final-state radiation is simulated using the {\footnotesize PHOTOS}~\cite{Barberio:1990ms} package for all charged final-state particles.

A list of the background MC samples is presented in Table~\ref{tab:3.1} with the ratio of the number of generated events $N^\text{gen}_\text{MC}$ to the expected number of corresponding events in data (product of the integrated luminosity $\mathcal{L}^\text{int}_\text{data}$ and the process cross section $\sigma_\text{proc}$).
\begin{table}[htb]
\caption{ Background MC samples with their size.} \label{tab:3.1}
\begin{tabular}
 {@{\hspace{0.5cm}}l@{\hspace{0.5cm}}  @{\hspace{0.4cm}}c@{\hspace{0.5cm}}}
\hline \hline
Processes & $N^\text{gen}_\text{MC}/(\mathcal{L}^\text{int}_\text{data}\sigma_\text{proc})$ \\
\hline
$\ee\to\tat$ background & 4.5 \\
$\ee\to\mam$ & 4.4 \\
$\ee\to\qq$ ($q=u,\,d,\,s$, $c$) & 5.8 \\
$\ee\to\Upsilon(4S)\to B\bar{B}$ & 10.2 \\
$\ee\to\ee\lal$ ($\ell=e,\,\mu$) & 6.9 \\
$\ee\to\ee\qq$ ($q=u,\,d$) & 7.5 \\
$\ee\to\ee\qq$ ($q=s,\,c$) & 8.1 \\
Bhabha scattering & 0.2\\
\hline \hline
\end{tabular}
\end{table}

\section{Event selection}

The event selection is performed in three steps. The first step is the preselection of candidates in $\tat$ events with the $\tat$-pair decay topology of interest. The second step is dedicated to the kink candidate selection. In the last step, we apply the BDT (boosted decision tree classifier) machine learning algorithm~\cite{FREUND1997119,Hocker:2007ht} to select signal event candidates and suppress the kink background. 

\subsection{Preselection}\label{sec:4.1}

In the first step, $\tat$-pair event candidates are required to pass the preliminary selection criteria. They are used to select the $\tat$-pair decay topology and suppress the contribution from Bhabha scattering, $\ee\to\mam$, two-photon production, $\ee\to q\bar q$ ($q=u$, $d$, $s$, or $c$), and $B\bar{B}$ events.

In the c.m. frame, the event is divided by the plane perpendicular to the thrust vector $\vec{n}_T$ into two hemispheres. The vector $\vec{n}_T$ is defined as follows
\begin{eqnarray}\label{eq:4.1}
T=\max_{\vec n_T}\frac{\sum_i|\vec p_i\cdot \vec n_T|}{\sum_i|\vec{p}_i|}.
\end{eqnarray} 
Here $\vec{p}_i$ is the momentum of the $i$th track; the summation is over all tracks in the event. The signal hemisphere is determined by the muon candidate momentum direction. The complementary one is called tagging hemisphere.

In the present analysis, the decay mode of the second $\tau$ lepton is not important. Therefore, our selection includes only the information about the event topology formed by charged tracks from the IP. In the signal hemisphere, we require only one track from the IP with the impact parameters in the $r\phi$-plane and along the $z$-axis (the direction opposite to the $e^+$ beam) to be $dr_\text{sig}<2\,\text{cm}$ and $|dz_\text{sig}|<4\,\text{cm}$, respectively. We also require one secondary electron candidate track in the signal hemisphere; however, at this step, the parameters of this track are not used. As the $\tau$ lepton decays dominantly into one or three charged tracks in the final state, in the tagging hemisphere, we require one (topology 1--1) or three (topology 1--3) charged tracks from the IP with their impact parameters to be $dr_\text{tag}<0.5\,\text{cm}$ and $|dz_\text{tag}|<2\,\text{cm}$. The total charge of the event is required to be zero.

Some events may contain photons, for example, from $\pi^0$s. They are selected with the energy requirement $E_\gamma>50\,\text{MeV}$. In the signal hemisphere, the maximum photon energy and the total sum of the photon energies are limited to be less than $300\,\mev$ and $400\,\mev$, respectively, and the $\pi^0$ candidates (a combination of two photons with $|M(\gamma\gamma)-m_{\pi^0}|<15\,\mevcc$, corresponding to approximately $\pm3\sigma$ window in the resolution) are vetoed.

For topology 1--1, the primary backgrounds are Bhabha scattering, two-photon interactions, $\ee\to\mam$, and $\ee\to q\bar q$ ($q=u$, $d$, $s$, or $c$). For topology 1--3, the main background is $\ee\to q\bar q$ ($q=u$, $d$, $s$, or $c$). To suppress the contribution of these processes, additional requirements are used. They are based on the fact that the $\ee\to\tat$ events with the $\tat$-pair subsequent decay are characterized by a large missing energy ($E_\text{miss}$) and missing momentum ($\vec{p}_\text{miss}$) due to undetected neutrinos. The missing four-momentum $(\vec{p}_\text{miss},\,E_\text{miss})$ is defined as follows
\begin{eqnarray}\label{eq:4.2}
P_\text{miss}=P^{*}_\text{beam}-P^{*}_\text{trk(IP)}-P^{*}_\gamma,
\end{eqnarray} 
where $P^{*}_\text{beam}$ is the beam four-momentum in the c.m. frame, $P^{*}_\text{trk(IP)}$ is a sum of four-momenta of all tracks from the $\tat$-pair in the c.m. frame, and $P^{*}_\gamma$ is a sum of four-momenta of all photons in the c.m. frame. Another feature of $\tat$ events is a nearly uniform distribution of $\cos{\theta_{\text{miss}}}$, where $\theta_{\text{miss}}$ is the angle between the $\vec{p}_\text{miss}$ and $z$-axis.

Thus, we apply the requirements on the missing mass ($m_\text{miss}^2=P_\text{miss}^2$) $1\,\gevcc<m_\text{miss}<7\,\gevcc$, missing angle $\pi/6<\theta_{\text{miss}}<5\pi/6$, thrust magnitude $0.85<T<0.99$, and invariant mass of the tag-side tracks $m^\text{tag}_\text{trk}<1.8\,\gevcc$.

To suppress the remaining Bhabha scattering contribution, we apply an electron veto for the tag-side track for topology 1--1 using identification based on the information from the CDC, ACC, and ECL~\cite{Nakano:2002jw}. We require its likelihood ratio $\mathcal{R}(e/x)=\mathcal{L}_e/(\mathcal{L}_e+\mathcal{L}_x)$ to be less than $0.4$, where $\mathcal{L}_e$ and $\mathcal{L}_x$ are the likelihood values of the track for the electron and non-electron hypotheses, respectively. This requirement rejects about 80\% of events with an electron on the tag side.

\subsection{Kink selection}\label{sec:4.2}

In this subsection, we describe the preselection of candidates for events with the $\menn$ decay in the CDC. As mentioned above, the daughter electron track originating from the muon decay in the signal hemisphere is required to infer the muon polarization. To suppress random combinations with tracks from IP, we require the electron candidate impact parameter in the $r\phi$-plane to be $dr_e>4\,\text{cm}$. To reconstruct the $\menn$ decay inside the CDC, both the muon track and the electron track have to be reconstructed, leaving enough hits in the tracker. The last point of the muon track and the first point of the electron track must be inside the CDC, detached at least $10\,\text{cm}$ from its walls.

The track helix is parametrized by five parameters, whose determination requires at least five hits in the CDC. It is also important to discard fake tracks; thus, it is required for the total number of the CDC hits to be larger than $7$ for the electron candidates and larger than $10$ for the muon candidates. Both tracks from the $\menn$ decay are shorter than the average track of the nondecayed particle from IP; therefore, we require the number of their CDC hits to be less than $40$. 
Since the decayed muon does not leave the drift chamber, the absence of associated hits in the outer TOF, ECL, and KLM systems is required. The electron tracks originate outside the SVD and are stopped in the ECL; therefore, for them, we veto signals from the SVD and KLM systems.

Finally, the distance between the muon and electron tracks at the decay vertex is required to be less than $5\,\text{cm}$. This requirement is loose enough to keep almost 100\% of kink events while rejecting random combinations of tracks.

The overwhelming majority of events that passed these selection criteria have the form of a track kink. One of these processes is $\menn$, and the rest are backgrounds, which mimic the signal. They are light meson decays ($\pimunu$, $\kmunu$, $\kpmunu$, $\kpenu$, $\kpipi$, $\kcpi$, $\kpipipi$), and electron scattering, muon scattering, and hadron scattering. In Table~\ref{tab:4.1}, the signal and the main background processes are listed with their relative contributions. About $20\%$ of pion decay events, $30\%$ of kaon decay events, and $30\%$ of hadron scattering events come from $\ee\to q\bar q$, while all other events are mainly from $\ee\to\tat$.
\begin{table}[htb]
\caption{Relative contribution of the signal and background processes after the kink selection and before applying the BDT
requirement.} \label{tab:4.1}
\begin{tabular}
 {@{\hspace{0.5cm}}l@{\hspace{0.5cm}}  @{\hspace{0.5cm}}c@{\hspace{0.5cm}}}
\hline \hline
Type & Contribution (\%) \\
\hline
$\menn$ & 3.2 \\
$\pimunu$ & 22.4 \\
$\kpipi$ & 3.3 \\
$K^-\to3$ body & 4.6 \\
$K^-\to\mu^-\bar{\nu}_\mu$ & 45.9 \\
$e$-scattering & 9.5 \\
$\mu$-scattering & 1.1 \\
hadron scattering & 10.0 \\
\hline \hline
\end{tabular}
\end{table}

The kinks, formed by a decay-in-flight, are characterized by daughter particle kinematics in the mother particle rest frame determined by the momentum magnitude and emission angle. These two variables are only defined for the correct pair of mass hypotheses assigned to the tracks, e.g., for $\menn$, they are electron and muon mass hypotheses assigned to the daughter and mother particles, respectively. To indicate which pair is used in the particular case, we introduce the following notation: $p_{p_1p_2}$ and $\theta_{p_1p_2}$ mean the daughter particle momentum and emission angle in the mother particle rest frame with $p_1$ and $p_2$ mass hypotheses assigned to the daughter and mother tracks, respectively. Here we measure the daughter particle emission angle from the direction of the mother particle in the laboratory frame because this angle determines the efficiency to reconstruct a decay-in-flight. The efficiency to reconstruct the daughter track from a kink has a maximum for the daughter particles emitted perpendicular to the mother particle direction, while it drops for daughter particle emitted along the muon direction.

\begin{figure*}[!htbp]
\includegraphics[width=0.83\linewidth]{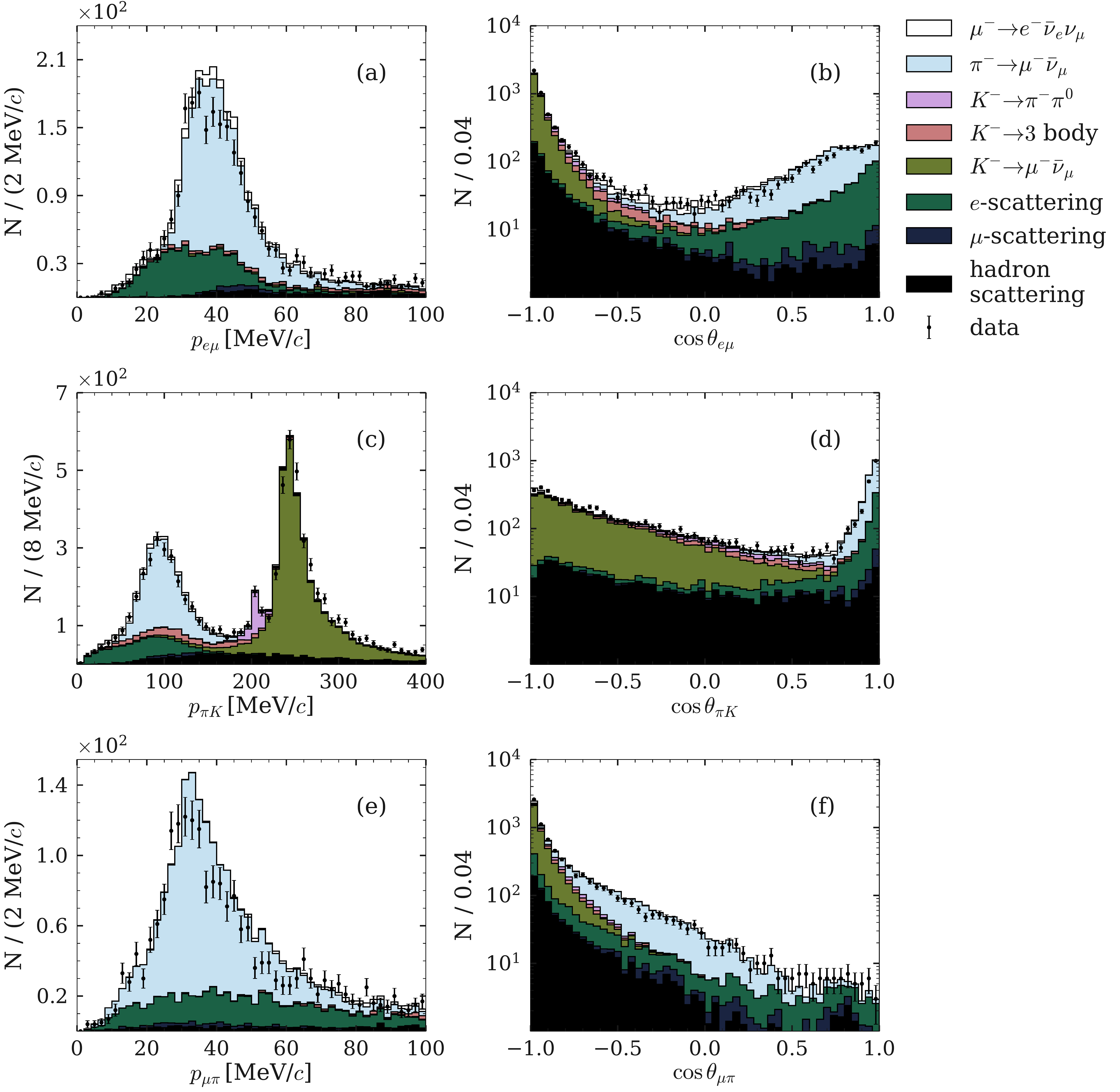}
\caption{Momentum $p_{p_1p_2}$ and angular $\cos{\theta_{p_1p_2}}$ distributions for the daughter particle (the mass hypothesis $p_1$) in the mother particle (the mass hypothesis $p_2$) rest frame. (a) $p_{e\mu}$, (b) $\cos{\theta_{e\mu}}$, (c) $p_{\pi K}$, (d) $\cos{\theta_{\pi K}}$, (e) $p_{\mu\pi}$, and (f) $\cos{\theta_{\mu\pi}}$.}
\label{fig:4.1}
\end{figure*}

In the present study, we use three pairs $(p_1,\,p_2)$: $(e,\,\mu)$, $(\pi,\,K)$, and $(\mu,\,\pi)$. For these mass hypotheses, we plot $p_{p_1p_2}$ and $\cos{\theta_{p_1p_2}}$ distributions in Fig.~\ref{fig:4.1}. A good agreement between the MC simulation (filled histograms) and the data (points with errors) is observed.

Both Table~\ref{tab:4.1} and Fig.~\ref{fig:4.1} show that the largest contribution to the background comes from the pion and kaon two-body decays. These processes are characterized by a peak in the momentum distribution of the daughter particle in the rest frame of the decayed one, which is clearly observed for the $\kmunu$ and $\kpipi$ decays in Fig.~\ref{fig:4.1}(c) and for the $\pimunu$ decay in Fig.~\ref{fig:4.1}(e), proving the correctness of the applied kink selection procedure. Before the selection based on the BDT, the signal contribution is small and hardly visible in Fig.~\ref{fig:4.1}. The signal shape has no sharp structures due to a three-body decay, and it is further smeared in the variables calculated with the wrong pair of mass hypotheses.

We define the signal region for $p_{e\mu}<70\,\mevc$. As can be seen from Fig.~\ref{fig:4.1}(a), the largest background contribution to this region is from the pion decay and electron scattering. However, in the region for the $\pimunu$ decay, $p_{\mu\pi}<100\,\mevc$, there are almost no $\menn$ events [see Fig.~\ref{fig:4.1}(e)], which makes it possible to effectively suppress the $\pimunu$ background, as well as a significant part of the electron scattering events.

\begin{figure}[!htbp]
  \centering
  \includegraphics[width=1\linewidth]{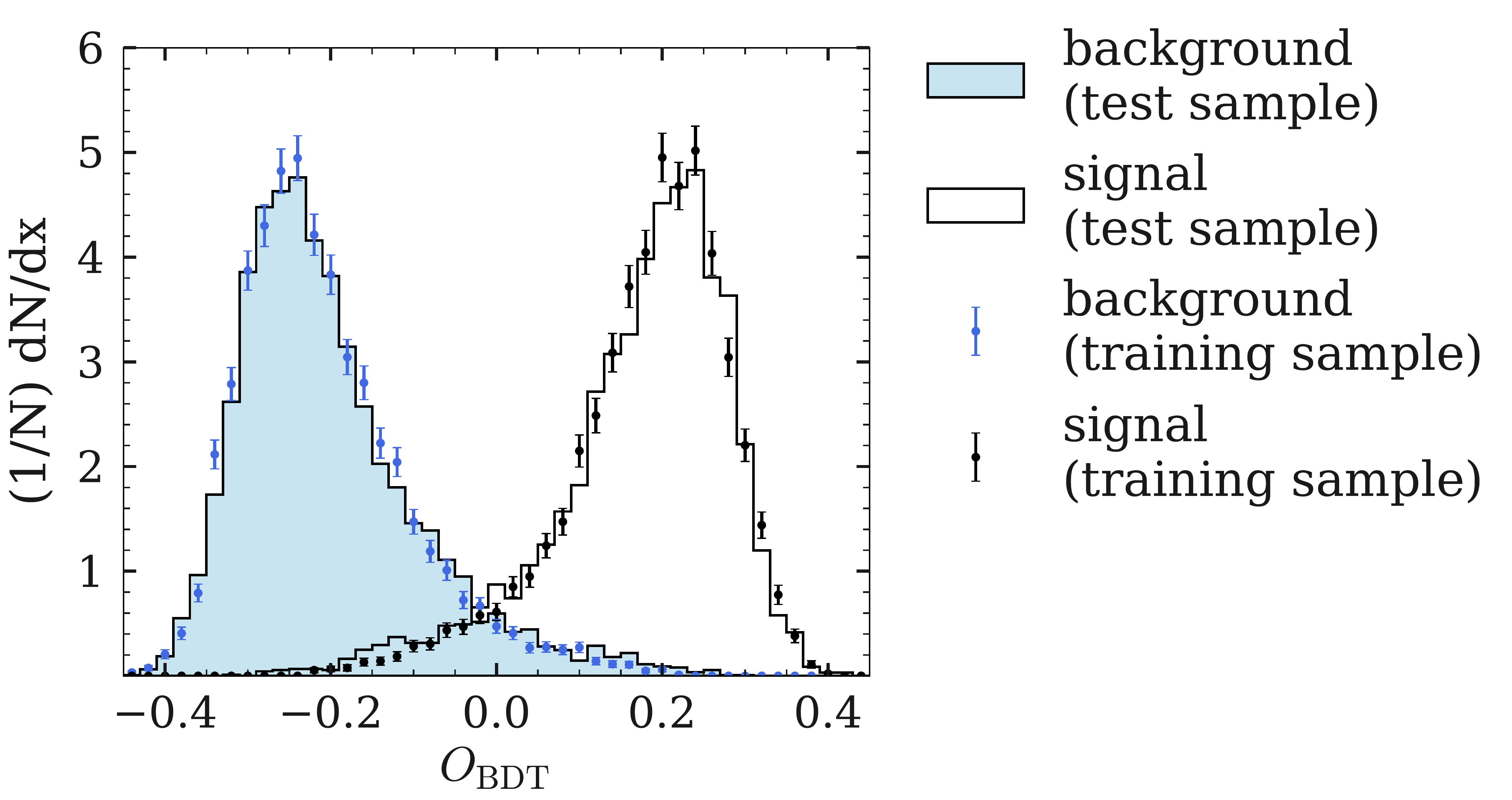}
\caption{$O_\text{BDT}$ distribution for the signal and background (training and test samples).}\label{fig:4.3}
\end{figure}

\subsection{BDT based signal selection}
To further suppress the background, we apply the BDT machine learning (ML) classification algorithm. To separate the signal from the background, we select twelve features based on the physics of the background processes. The first two features are $p_{\mu\pi}$ and $p_{\pi K}$. The next group of five features is responsible for the particle identification (PID) of muon and electron candidates. They are defined as likelihood ratios $\mathcal{R}(\ell/x)=\mathcal{L}_{\ell}/(\mathcal{L}_{\ell}+\mathcal{L}_x)$, where $\ell = \mu$ or $e$, and $\mathcal{L}_{\ell}$ and $\mathcal{L}_x$ are the likelihood values of the track for the muon (electron) and non-muon (non-electron) hypotheses, respectively. For muon candidates, we use PID based on the $dE/dx$ losses inside the CDC against electron, pion, kaon, and proton hypotheses ($x=e$, $\pi$, $K$, and $p$, respectively). For electron candidates, PID is based on the $dE/dx$ losses inside the CDC and ECL information; here only $\mathcal{R}(e/\mu)$ is used. Two more features are related to the decay vertex; they are the $z$-coordinate of the last point of the mother particle track and the distance between the daughter and mother tracks at the decay vertex. Finally, to suppress the residual contribution from $\ee\to \qq$, we use $m_\text{miss}$, $\cos{\theta_\text{miss}}$, and thrust magnitude as separation variables.

Although the $\cos{\theta_{p_1p_2}}$ variables show a good separation power (see Fig.~\ref{fig:4.1}), we do not use them in the BDT because they are strongly correlated with $\cos{\theta_{e}}$ (the main variable to fit $\xip$) and, therefore, bias the $\xip$ measurement with poorly controlled systematics.

The distribution of the BDT output variable $O_\text{BDT}$ is shown in Fig.~\ref{fig:4.3} for signal and background for training and test samples. The optimal selection of $O_\text{BDT}>0.0979$ is obtained by maximizing the ratio $N_\text{sig}/\sqrt{N_\text{sig}+N_\text{bckg}}$, where $N_\text{sig}$ is the number of selected signal events, and $N_\text{bckg}$ is the number of selected background events. The obtained signal selection efficiency is $\varepsilon_\text{sig}\approx80\%$, while the background is suppressed by a factor of fifty.

\begin{figure}[!htbp]
  \includegraphics[width=1.0\linewidth]{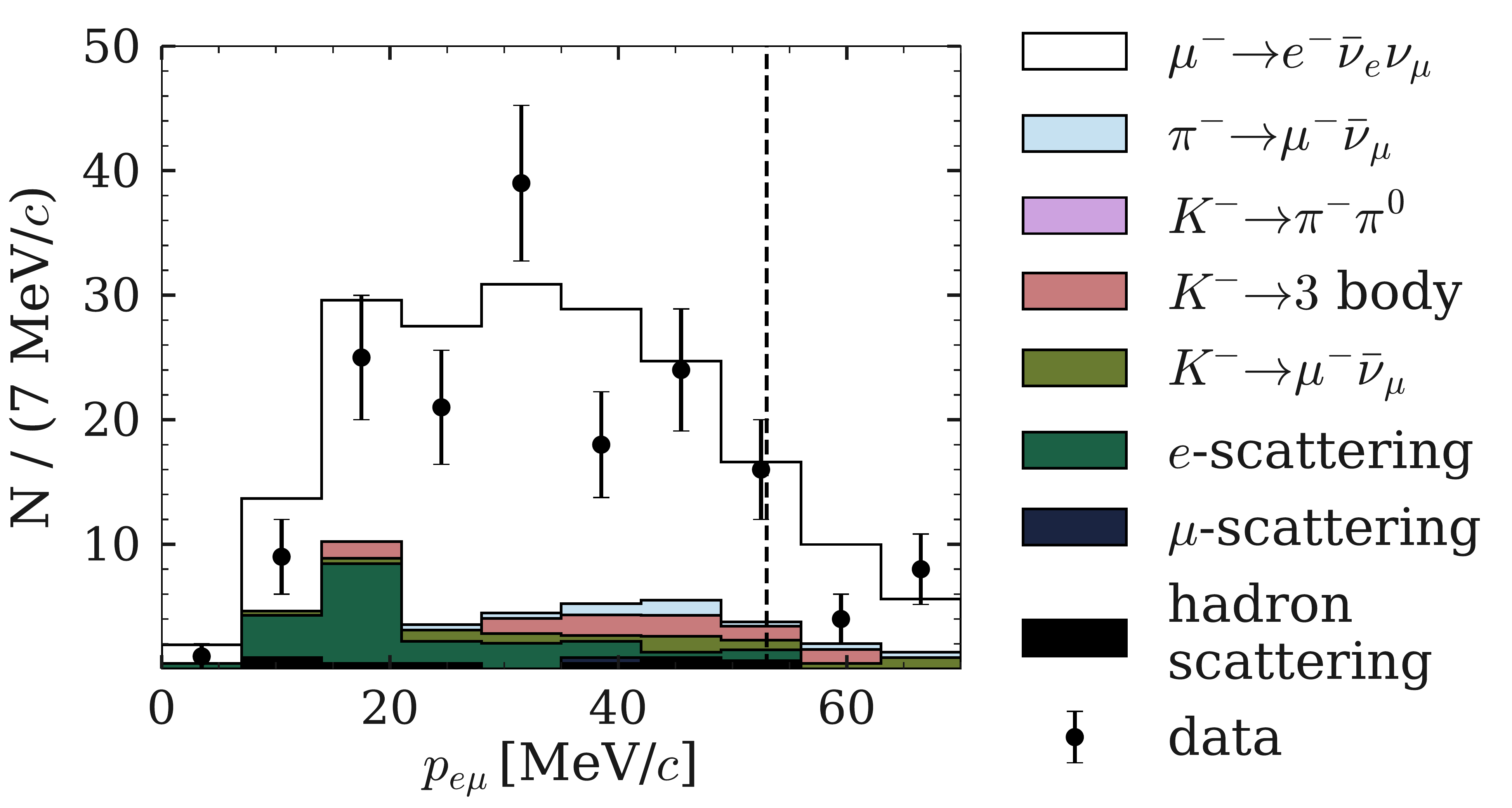}
\caption{Momentum distribution for the electron and muon mass hypotheses for the daughter and mother particles, respectively. The dashed line shows the $53\,\text{MeV}/c$ threshold.
}
\label{fig:4.4}
\end{figure}

To illustrate the performance of BDT, we plot the electron candidate momentum in the muon rest frame shown in Fig.~\ref{fig:4.4}. 
The absence of the Belle track reconstruction algorithm optimization for the kink events leads to a wide tail above the kinematic threshold of $53\,\text{MeV}/c$ in the $\menn$ decay. The relative contribution of the signal and background processes after the BDT application is listed in Table~\ref{tab:4.2}. 
About $6\%$ of the $\menn$ decays come from the non-$\tat$ events.

\begin{table}[!htb]
\caption{ Relative contribution of the signal and background processes after the BDT application.} \label{tab:4.2}
\begin{tabular}
 {@{\hspace{0.5cm}}l@{\hspace{0.5cm}}  @{\hspace{0.5cm}}c@{\hspace{0.5cm}}}
\hline \hline
Type & Contribution (\%) \\
\hline
$\menn$ & 77.8 \\
$\pimunu$ & 2.2 \\
$K^-\to3$ body & 4.3 \\
$K^-\to\mu^-\bar{\nu}_\mu$ & 3.7 \\
$e$-scattering & 9.6 \\
$\mu$-scattering & 0.2 \\
hadron scattering & 2.2 \\
\hline \hline
\end{tabular}
\end{table}
Finally, the number of the reconstructed signal $\menn$ decays and background events are estimated from the MC simulation to be $139\pm2$ and $50\pm5$, respectively, where the uncertainty is due to the limited size of the MC samples. In the data, $165$ signal-candidate events pass all the applied selection criteria.

\section{Background study}\label{sec:5}
In the present study, the background suppression and determination of the fit function are based on the MC simulation; thus, it is important to control the differences between the MC samples and the data and take them into account as systematic uncertainties. Therefore, we conduct a study of background processes in the data and the MC simulation using large pure samples with different types of kink candidates (pion and kaon decays, hadron and electron scattering).

Light meson decays are selected in two ways. The first method is based on the BDT described in the previous section, where we mark $\pimunu$ or $\kmunu$ decay as a signal. The samples obtained in this way have a purity close to unity. 

\begin{figure}[htb]
  \includegraphics[width=1\linewidth]{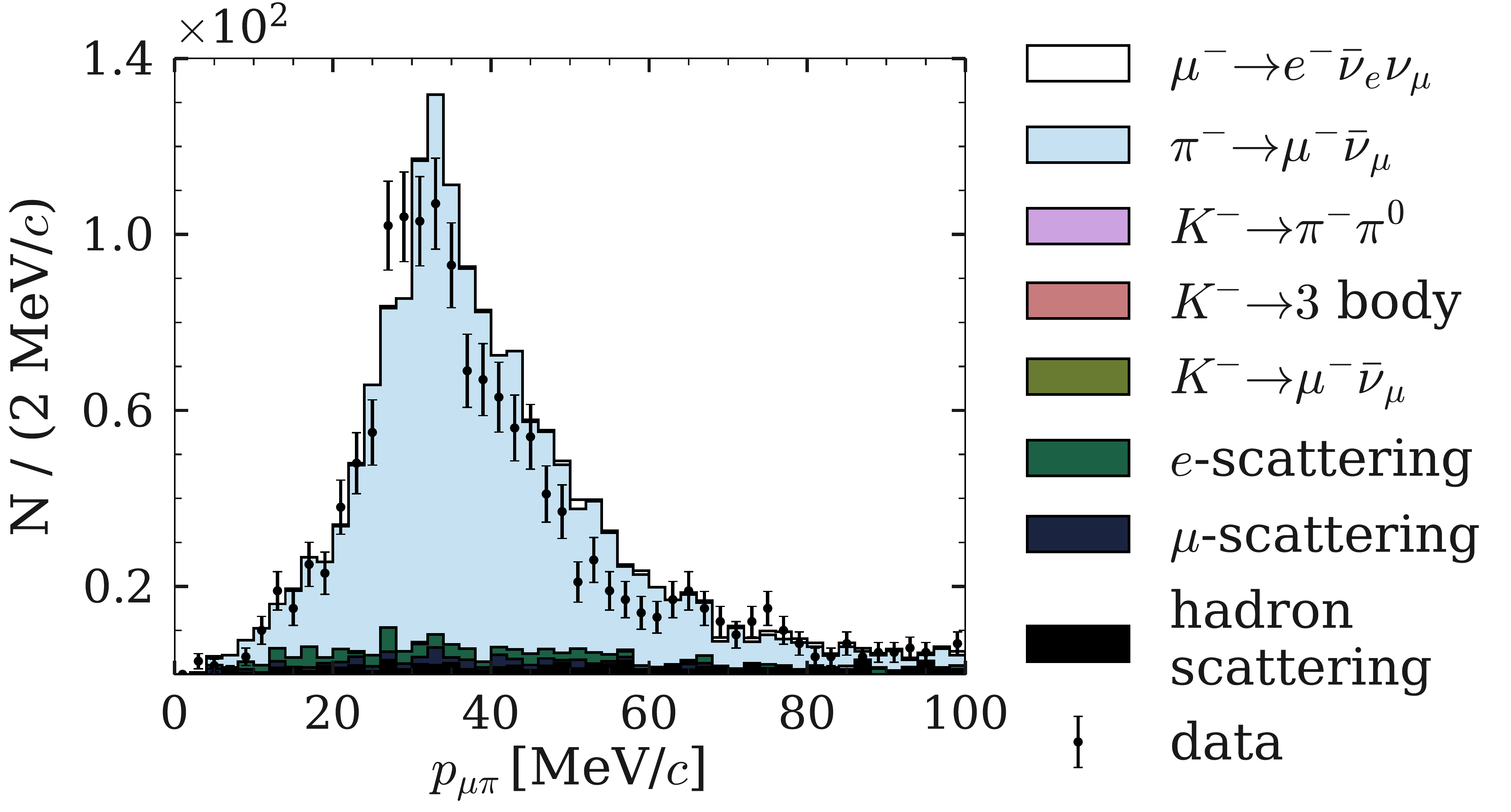}
\caption{$p_{\mu\pi}$ distribution for the $\pimunu$ event candidates selected with the BDT from the $\tau$ sample.}
\label{fig:5.1}
\end{figure}

\begin{figure}[htb]
  \includegraphics[width=1\linewidth]{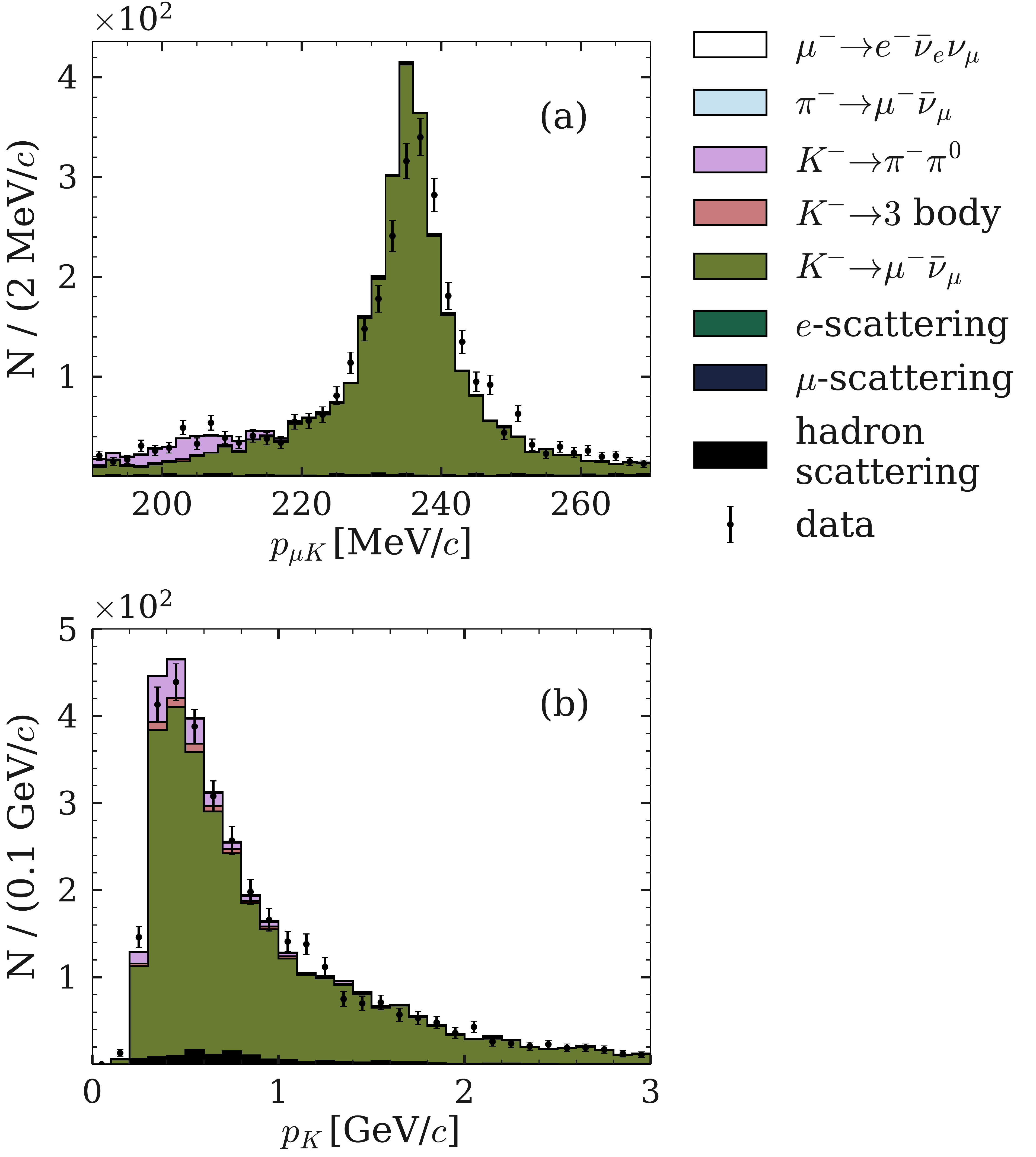}
\caption{(a) $p_{\mu K}$ and (b) $p_K$ distributions for the $\kmunu$ event candidates selected with the BDT from the $\tau$ sample.}
\label{fig:5.2}
\end{figure}

The distributions of $p_{\mu\pi}$ for the selected $\pimunu$ sample and $p_{\mu K}$ for the selected $\kmunu$ sample are shown in Fig.~\ref{fig:5.1} and Fig.~\ref{fig:5.2}(a), respectively.
In the former plot, the $p_{\mu\pi}$ distribution in the MC sample is shifted to the higher muon momentum compared to the data. 
This effect is related to the imperfection of the track reconstruction algorithm in the case of a kink and is especially pronounced in the pion decay due to the low energy release. In Fig.~\ref{fig:5.2}(a), we observe that the muon momentum peak has a larger width in the data indicating a better resolution in the MC simulation, while kaon momentum distribution in the laboratory frame $p_K$ plotted in Fig.~\ref{fig:5.2}(b) shows an agreement between the MC simulation and the data within statistical uncertainties of both samples.

\begin{figure*}[!htbp]
\includegraphics[width=0.83\linewidth]{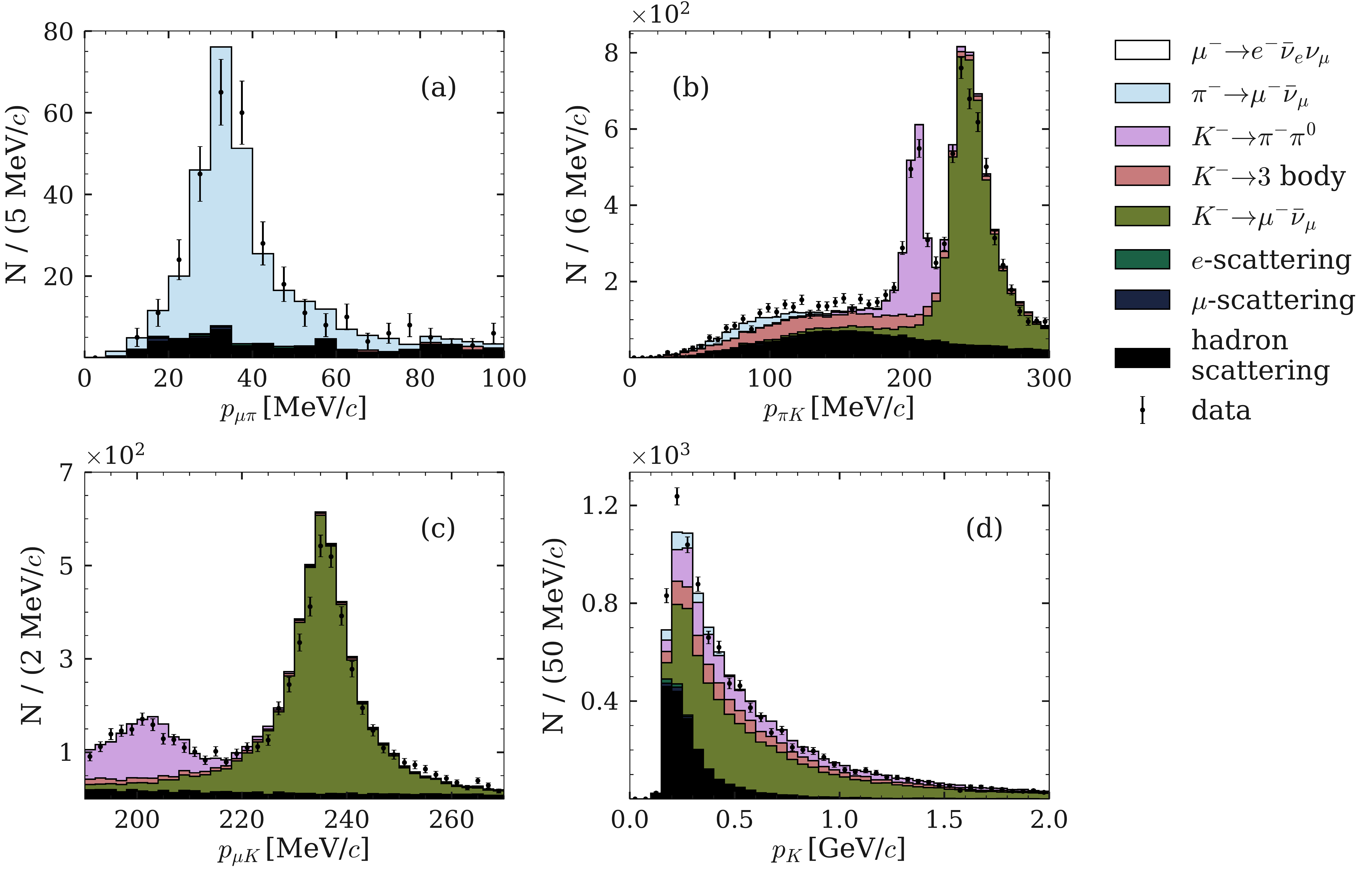}
\caption{$D^{*+}$ sample. (a) $p_{\mu\pi}$ distribution for pion kinks; (b) $p_{\pi K}$, (c) $p_{\mu K}$, and (d) $p_K$ distributions for kaon kinks.}
\label{fig:5.3}
\end{figure*}

It is also important to control systematics caused by the BDT application; thus, background processes have to be studied in samples obtained without ML algorithms. To obtain a high-purity kink sample without BDT, we use the decay chain $D^{*+}\to D^0(\to K^-\pi^+)\pi^+$ since there is a large sample of $D^{*+}$ mesons collected by the Belle detector, and both decays, $D^{*+}\to D^0\pi^+$ and $D^0\to K^-\pi^+$, are well-studied. In these events, it is possible to reconstruct the light meson decay-in-flight and tag the kink type. A detailed description of the $D^{*+}$ sample selection is given in Appendix~\ref{app:1}. 

For the $\pimunu$ kinks selected from the $D^{*+}$ sample, we plot the $p_{\mu\pi}$ distribution in Fig.~\ref{fig:5.3}(a). It is similar to Fig.~\ref{fig:5.1}, although the statistics are several times smaller.

Concerning kaon kinks from the $D^{*+}$ sample, they include both kaon two-body and three-body decays and a large number of events with hadron scattering. To illustrate the abundance of selected processes, we plot the $p_{\pi K}$ distribution in Fig.~\ref{fig:5.3}(b).
Here we observe a relatively large contribution of $\kpipi$ decays compared to Fig.~\ref{fig:4.1}(c), where such events are suppressed by requirements for photons and $\pi^0$s. For this kaon decay mode, we confirm the agreement between the MC simulation and the data within statistical uncertainties. For $\kmunu$ decays, we observe a discrepancy; therefore, to study this process in more detail, we plot the $p_{\mu K}$ distribution in Fig.~\ref{fig:5.3}(c).
Here the discrepancy in the muon momentum peak width for $\kmunu$ decays is observed to be similar to one in Fig.~\ref{fig:5.2}(a) for the $\tau$ sample and thus confirms this to be a systematic effect.

Kaon kinks from the $D^{*+}$ sample also include hadron scattering events [e.g., Fig.~\ref{fig:5.3}(b)]. This process is typical for slow hadrons, as can be seen from Fig.~\ref{fig:5.3}(d), where $p_K$ is plotted. 
For the first two bins with data, we observe a significant discrepancy between the data and MC samples, while an agreement is observed for kaon kinks from the $\tau$ sample [Fig.~\ref{fig:5.2}(b)]. Another confirmation that the MC simulation does not reproduce hadron scattering is an underestimation of the events number in the hadron scattering region observed in Fig.~\ref{fig:5.3}(b). The difference between the MC simulation and the data is expected since this process is not perfectly described by {\footnotesize GEANT3}. For larger $p_K$, the MC simulation reproduces the data within statistical uncertainties for both $D^{*+}$ and $\tau$ samples.

\begin{figure}[htb]
  \includegraphics[width=1\linewidth]{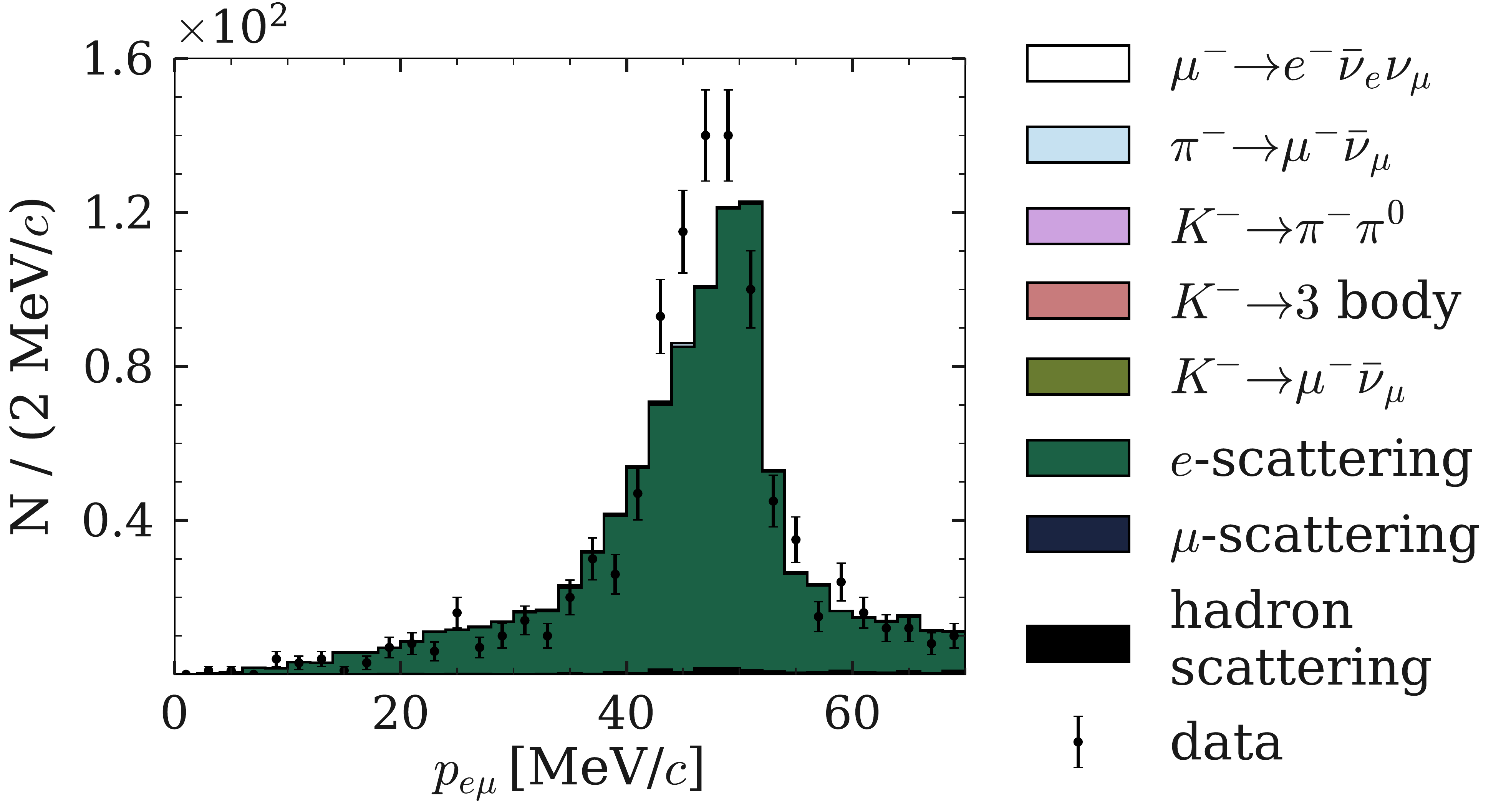}
\caption{$p_{e \mu}$ distribution for the electron scattering kinks selected from the $\gamma$-conversion sample.}
\label{fig:5.4}
\end{figure}

The electron scattering process makes a significant contribution to the background. The study of this process is based on the sample obtained from the $\gamma$-conversion on the detector material in the IP vicinity. The selection of the $\gamma$-conversion sample is described in detail in Appendix~\ref{app:2}. To illustrate the electron scattering process, we use the same pair of mass hypotheses as in the fit of the $\menn$ process. The $p_{e\mu}$ distribution is shown in Fig.~\ref{fig:5.4}.
A discrepancy between the MC and data samples is observed in the shape of the electron spectrum and taken into account in the systematics.

In conclusion, a complete study of the main background processes is done. All observed discrepancies are taken into account as systematic uncertainties. In addition, the discussed samples also provide important information about secondary and primary tracks in events that contain kinks for all main particle types except primary muon. This information is also used in systematics estimation, as it is described in the corresponding section below.

\section{Fit function and fit result}

According to Eq.~\eqref{eq:2.1}, the term proportional to $\xip$ depends on $\cos{\theta_{e}}$, $y$, and $x$. Since the dependence on $x$ is very weak,\footnote{If $x_0^2$ is neglected in Eq.~\eqref{eq:2.1}, the dependence on $x$ factorizes.} we can integrate over it without loss of sensitivity. In contrast, the dependence on $y$ is strong, and integration over it dramatically decreases the sensitivity to $\xip$; thus, we perform a two-dimensional (2D) fit on the $(y,\,\cos{\theta_{e}})\equiv(y,\,c)$ distribution using an unbinned maximum-likelihood method. 

The likelihood function is
\begin{eqnarray}\label{eq:6.1}
    \mathcal{L}=\prod\limits_{i=1}^{n}\mathcal{P}(y_i,\,c_i;\, \xip),
\end{eqnarray}
where $n$ denotes the number of data events, and $\mathcal{P}(y,\,c;\, \xip)$ is a probability density function (PDF)
\begin{eqnarray}\label{eq:6.2}
\begin{aligned}
    \mathcal{P}(y,\,c;\, \xip)=p\,\mathcal{P}_\text{sig}(y,\,c;\, \xip) + (1-p)\, \mathcal{P}_\text{bckg}(y,\,c).
\end{aligned}
\end{eqnarray}
Here $p=N_\text{sig}/(N_\text{sig}+N_\text{bckg})=0.74$ is the signal purity, the ratio of the number of signal events $N_\text{sig}$ to the total number of events $N_\text{sig}+N_\text{bckg}$, $\mathcal{P}_\text{sig}(y,\,c;\, \xip)$ is a signal PDF, and $\mathcal{P}_\text{bckg}(y,\,c)$ is a background PDF. The signal purity and both PDFs are obtained from the MC simulation.

The signal PDF can be determined from the theoretical PDF $\mathcal{P}_\text{th}(\tilde{y},\,\tilde{c},\,\boldsymbol{z};\, \xip)$ by applying efficiency corrections and performing a convolution with the detector resolution:
\begin{eqnarray}\label{eq:6.3}
\begin{aligned}
    \mathcal{P}_\text{sig}(y,\,c;\,\xip)&=
    \dfrac{1}{N(\xip)}
    \int \mathcal{P}_\text{th}(\tilde{y},\,\tilde{c},\,\boldsymbol{z};\, \xip)\\
    &\times\eta(\tilde{y},\,\tilde{c},\,\boldsymbol{z})\,
    g(y,\,c,\,\tilde{y},\,\tilde{c},\,\boldsymbol{z})\, d\boldsymbol{z}\, d\tilde{y}\, d{\tilde{c}},\\
    N(\xip)& =\int \mathcal{P}_\text{th}(\tilde{y},\,\tilde{c},\,\boldsymbol{z};\, \xip)\,\eta(\tilde{y},\,\tilde{c},\,\boldsymbol{z})\\
    &\times g(y,\,c,\,\tilde{y},\,\tilde{c},\,\boldsymbol{z})\,
    d\boldsymbol{z}\, d\tilde{y}\, d{\tilde{c}}\,dy\,dc.
    \end{aligned}
\end{eqnarray}
Here $(\tilde{y},\,\tilde{c})\equiv(\tilde{y},\,\cos{\tilde{\theta}_{e}})$ are ``true'' physical quantities of our interest, and $\boldsymbol{z}$ is a vector of the rest of the true physical variables not used in the fit. Functions $\eta(\tilde{y},\,\tilde{c},\,\boldsymbol{z})$ and $g(y,\,c,\,\tilde{y},\,\tilde{c},\,\boldsymbol{z})$ are the efficiency and resolution functions, respectively. The theoretical PDF, $\mathcal{P}_\text{th}(\tilde{y},\,\tilde{c},\,\boldsymbol{z};\, \xip)$, is obtained from the differential decay width given by Eq.~\eqref{eq:2.1}. 

Both efficiency and resolution functions are too complicated to express in analytic form; thus, it is almost impossible to calculate the $\mathcal{P}_\text{sig}(y,\,c;\,\xip)$ function given in Eq.~\eqref{eq:6.3} analytically. Fortunately, the theoretical PDF is linear in $\xip$; therefore, we rewrite it as follows
\begin{eqnarray}\label{eq:6.4}
    \mathcal{P}_\text{th}(\tilde{y},\,\tilde{c},\,\boldsymbol{z};\, \xip)=A(\tilde{y},\,\tilde{c},\,\boldsymbol{z})+\xip B(\tilde{y},\,\tilde{c},\,\boldsymbol{z}).
\end{eqnarray}
Using this form, we rewrite Eq.~\eqref{eq:6.3}:
\begin{eqnarray}\label{eq:6.5}
    \mathcal{P}_\text{sig}(y,\,c;\,\xip)=\dfrac{\bar{A}(y,\,c)+\xip\bar{B}(y,\,c)}{\tilde{A}+\xip\tilde{B}},
\end{eqnarray}
where
\begin{eqnarray}\label{eq:6.6}
\begin{aligned}
    \bar{A}(y,\,c)&=\int A(\tilde{y},\,\tilde{c},\,\boldsymbol{z})\,\eta(\tilde{y},\,\tilde{c},\,\boldsymbol{z})\\ 
    &\qquad\qquad\times g(y,\,c,\,\tilde{y},\,\tilde{c},\,\boldsymbol{z})\, d\boldsymbol{z}\,d\tilde{y}\,d\tilde{c},\\
    \bar{B}(y,\,c)&=\int B(\tilde{y},\,\tilde{c},\,\boldsymbol{z})\eta(\tilde{y},\,\tilde{c},\,\boldsymbol{z})\\
    &\qquad\qquad\times g(y,\,c,\,\tilde{y},\,\tilde{c},\,\boldsymbol{z})\, d\boldsymbol{z}\, d\tilde{y}\,d\tilde{c},\\
    \tilde{A}&=\int\bar{A}(y,\,c)\,dy\,dc,\quad\tilde{B}=\int\bar{B}(y,\,c)\,dy\,dc.
\end{aligned}
\end{eqnarray}
In this study, the dependence of the signal PDF normalization on $\xip$ is negligible as $\tilde{A}\gg\tilde{B}$. 

To calculate $\bar{A}(y,\,c)/\tilde{A}$ and $\bar{B}(y,\,c)/\tilde{A}$, we use two MC samples generated with $\xip=1$ and $\xip=-1$. 
Their distributions in $(y,\,c)$ are determined exactly by the following PDFs: $\mathcal{P}_{+1}(y,\,c)=\mathcal{P}_\text{sig}(y,\,c;\,+1)$ and $\mathcal{P}_{-1}(y,\,c)=\mathcal{P}_\text{sig}(y,\,c;\,-1)$, respectively, providing 
\begin{eqnarray}\label{eq:6.7}
    \mathcal{P}_\text{sig}(y,\,c;\,\xip)\!&=&\!\dfrac{1}{2}\left\{\mathcal{P}_{+1}(y,\,c) + \mathcal{P}_{-1}(y,\,c)\right.\nonumber\\
    &&\left.+\xip\left[\mathcal{P}_{+1}(y,\,c) - \mathcal{P}_{-1}(y,\,c)\right]\right\}.
\end{eqnarray}

All the PDFs can be obtained in the form of 2D histograms of $(y,\,c)$ or in the form of smooth functions describing the distributions in $(y,\,c)$. Since the signal MC sample statistics is large, 2D histograms of $(y,\,c)$ can already be considered as almost smooth functions (there are no statistically significant fluctuations), which can be used in the fit without loss of accuracy. Thus, for simplicity and naturalness, we obtain $\mathcal{P}_{\pm1}(y,\,c)$ in the form of the 2D histogram of $10\times10$ bins with an interpolation of the intermediate values. Alternatively, we use a smooth function to evaluate the systematic uncertainty as it is described in Sec.~\ref{sec:7.4}.

In contrast to the signal, the background MC sample has modest statistics, and there is no feasibility to increase it. Therefore, $\mathcal{P}_\text{bckg}(y,\,c)$ is obtained from the approximation of a $6\times6$-bin histogram of $(y,\,c)$ distribution by a smooth parametric function so that $\chi^2/\text{n.d.f.}\approx1$.

The fit procedure is tested on ensembles of 1000 statistically independent simulated samples of the size expected in the data with eleven $\xip$ seed values from $-1$ to 1 in steps of $0.2$, and no statistically significant biases are observed.

\begin{figure}[htb]
  \includegraphics[width=1\linewidth]{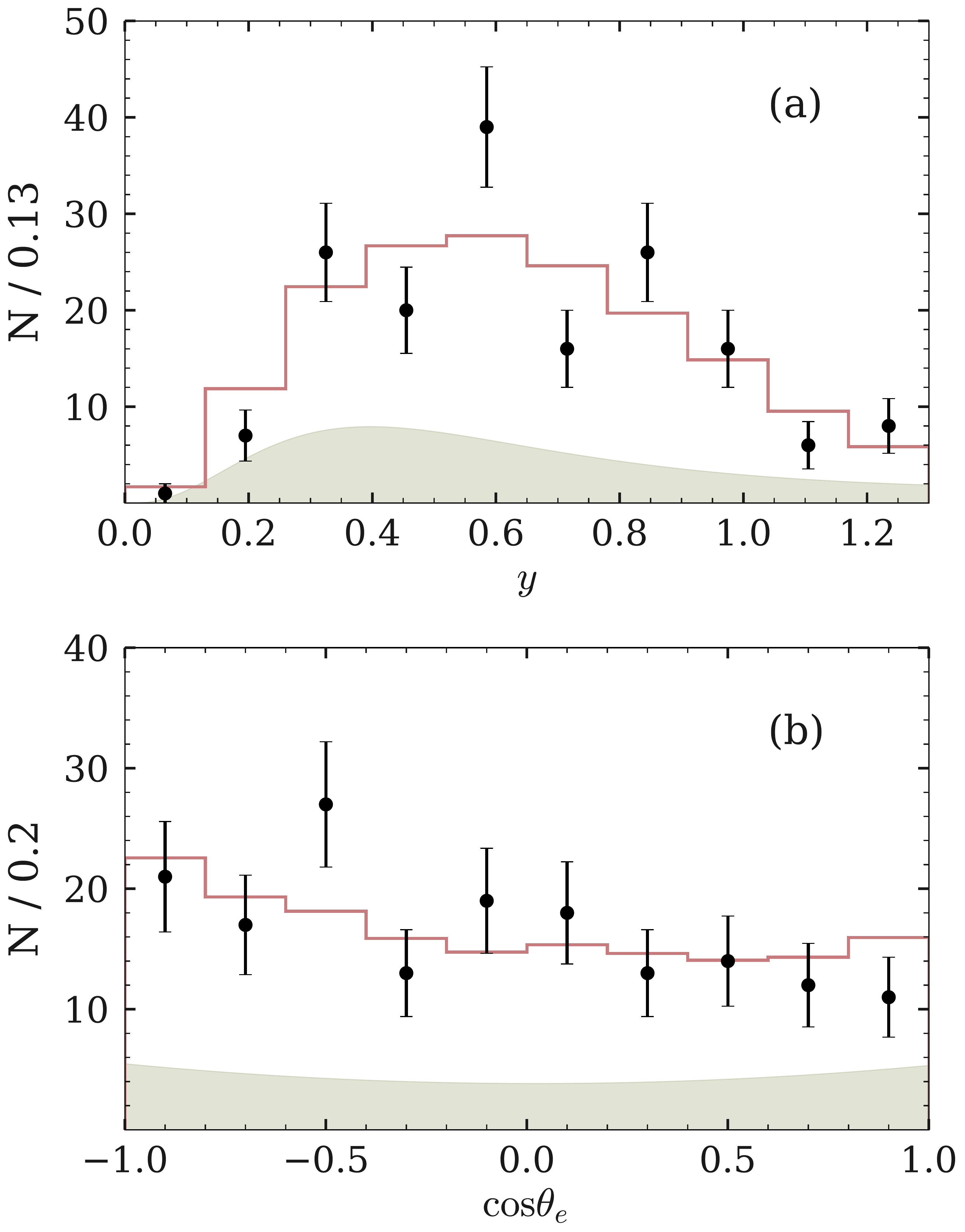}
\caption{The fit of the data with $\xip=0.22\pm0.94$. Points with errors correspond to the data, the solid line histogram corresponds to the fit function, and the shadowed area corresponds to the background function. (a)~projection onto $y$, (b)~projection onto $\cos{\theta_e}$.}
\label{fig:6.1}
\end{figure}

\begin{figure}[!htbp]
  \includegraphics[width=1\linewidth]{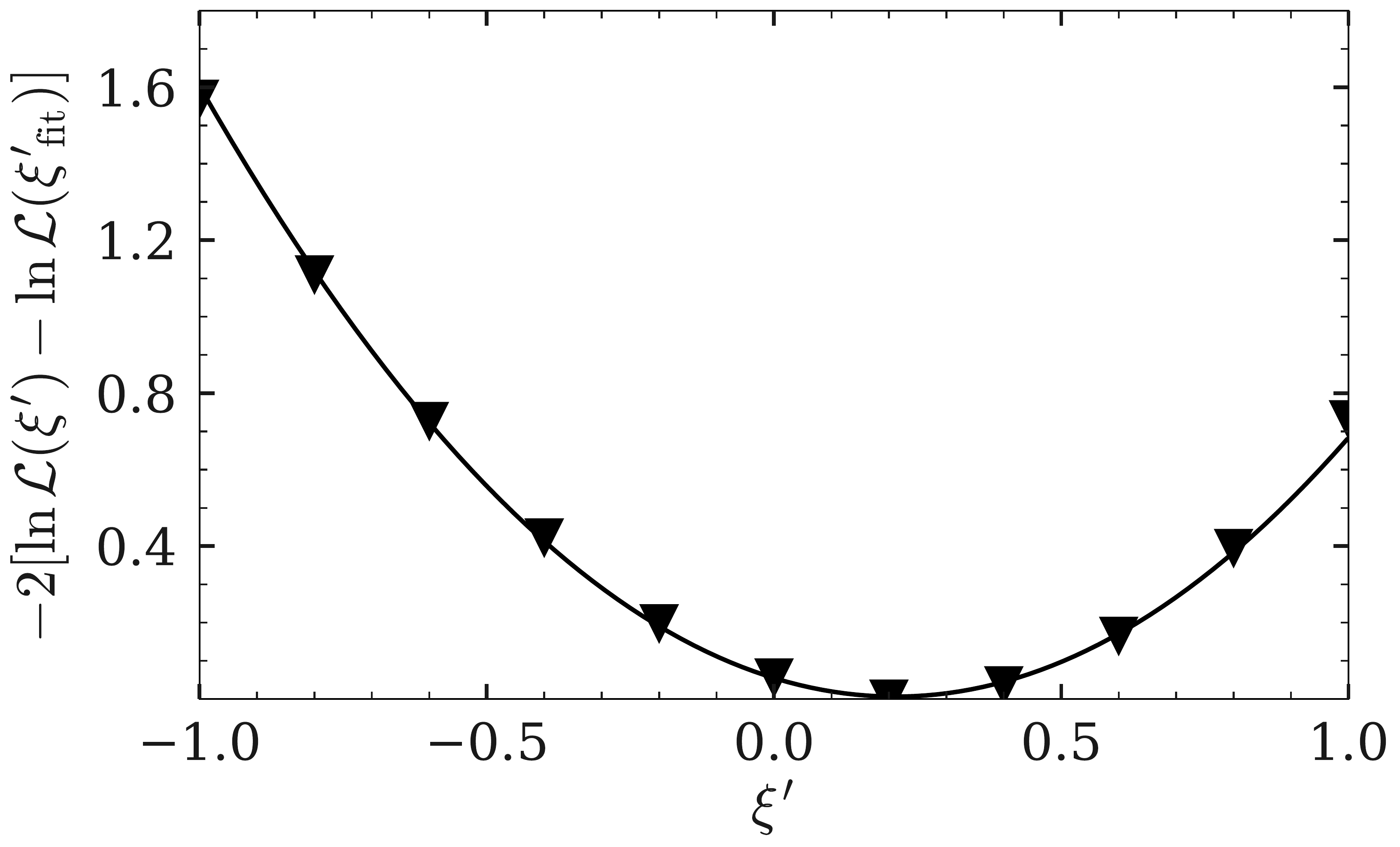}
\caption{The variation of the $-2\left[\ln{\mathcal{L}(\xip)} - \ln{\mathcal{L}(\xip_\text{fit})}\right]$ as a function of the $\xip$ value.}
\label{fig:6.2}
\end{figure}

Finally, the fit to the data yielded $\xip=0.22\pm0.94$. The projections of the data and the fit function onto the $y$ and $\cos{\theta_e}$ axes are shown in Fig.~\ref{fig:6.1}.
The variation of the $-2\left[\ln{\mathcal{L}(\xip)} - \ln{\mathcal{L}(\xip_\text{fit})}\right]$ as a function of the $\xip$ value is shown in Fig.~\ref{fig:6.2}. 
The $\xip=-1$ scenario is more than one standard deviation away from the measured $\xip$ value.

\begin{figure}[htb]
  \includegraphics[width=1\linewidth]{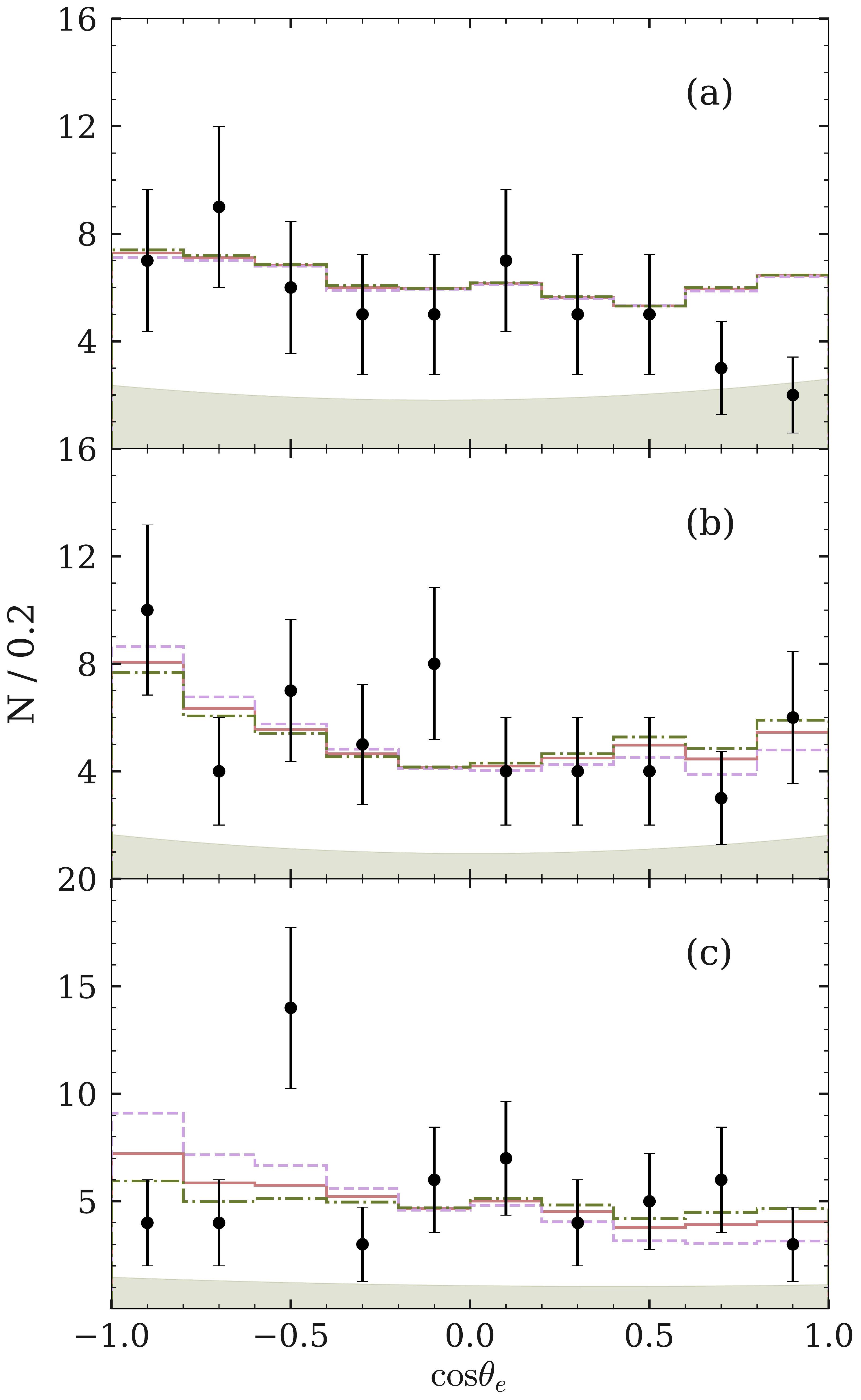}
\caption{Projection onto $\cos{\theta_e}$ for slices in $y$: (a)~$0 < y < 0.52$, (b)~$0.52 < y < 0.78$, and (c)~$0.78 < y < 1.3$. Points with errors correspond to the data, the solid line corresponds to the $\xip=0.22$ fit function, the dashed line corresponds to the $\xip=-1$ fit function, the dash-dotted line corresponds to the $\xip=1$ fit function, and the shadowed area corresponds to the background function.}
\label{fig:6.3}
\end{figure}

For a more detailed illustration of the fit, we plot three slices in $y$ ($0 < y < 0.52$, $0.52 < y < 0.78$, and $0.78 < y < 1.3$) projected onto $\cos{\theta_e}$ (Fig.~\ref{fig:6.3}). In addition to the fit function with $\xip=0.22$, we show fit functions with $\xip=-1$ (dashed) and with $\xip=1$ (dash-dotted).
For $y < 0.52$, there is almost no sensitivity to $\xip$, while for $0.78<y<1.3$, the sensitivity is maximum. This behavior is expected from the theoretical function given by Eq.~\eqref{eq:2.1}. The total $\chi^2$ for the fit projections shown in Fig.~\ref{fig:6.3} is $31$ with $\text{n.d.f.}=29$, demonstrating that the fit describes the data well. The total $\chi^2$ for the projections shown in Fig.~\ref{fig:6.3} for the function with $\xip=-1$ is $37$ and for the function with $\xip=1$ is $30$.

\section{Systematic uncertainties}
The systematic uncertainties are taken into account by assuming the most conservative approach. To estimate them, we generate for each source of the systematics an ensemble of 1000 toy MC samples with eleven $\xip$ seed values from $-1$ to 1 in steps of $0.2$ and the same statistics as estimated from the signal and background MC samples. Each sample is generated according to the 2D distribution in $y$ and $\cos{\theta_e}$ obtained from variation of the signal and background distributions within the expected uncertainties (observed discrepancies between the MC simulation and the data described in the previous sections). Then all samples are fitted with the default PDF function, and the average of obtained $\xip$ values over 1000 samples is calculated. The maximum difference between these mean values and the default one is taken as a systematic uncertainty. 

We also estimate the systematic uncertainties from the data by varying the PDF functions used in the fit to obtain the difference between a new $\xip$ value and the default one. We use this method as a crosscheck since it is less robust for systematics evaluation and always gives a lower value compared to the estimation from toy MC samples.

In this study, we distinguish four main categories of the systematic error sources: ``background,'' ``PID in BDT,'' ``signal,'' and ``fit procedure.''

\subsection{Background systematics}
This category of systematic errors includes the uncertainties in the expected background fraction of each type used in the fit PDF as well as the particular background shape. 

The signal purity $p$ is obtained from the MC simulation with the statistical uncertainty of $0.02$ induced by the limited number of signal and background MC events. While the signal MC sample is generated with large statistics, the size of the background MC sample is moderate, thus making a major contribution. The observed discrepancies between the data and simulation lead to an additional systematical uncertainty of $0.02$ in the purity value. 
The variation of $p$ within the combined error results in the systematic uncertainty of $0.10$. 

The PDF shape and relative contribution of each type of the background processes are the sources of the systematics because the MC simulation does not reproduce data perfectly. The main background contamination comes from the kaon and pion decays, electron and hadron scattering. For each of them, we conducted a small dedicated study described in Sec.~\ref{sec:5}. Prepared pure background samples allow us to observe discrepancies between the MC simulation and the data in both normalization and shape. To take these discrepancies into account, we reweight each type of the background MC sample based on particular kinematical characteristics. For the reweighted background sample, we obtain new values of the background smooth function parameters. The estimated systematic uncertainties are the following: $0.05$ for $\pimunu$, $0.06$ for $K^-\to 3$ body, $0.05$ for $\kmunu$, $0.10$ for $e$-scattering, and $0.11$ for hadron scattering. The statistical uncertainties of the parameters of the background PDF do not have much impact on the shape and have already been taken into account in the signal purity systematics.

The combined background systematic uncertainty is $0.20$.

\subsection{PID in BDT systematics}
To estimate the effects of PID usage in the BDT, we take advantage of the availability of various tagged kinks in the data selected without BDT application. This systematic uncertainty contains two separate contributions: PID uncertainties of primary muons and daughter electrons.

The PID uncertainty of the daughter track is easier to analyze since we have tagged secondary electrons from the electron scattering ($\gamma$-conversion sample), muons from kaon decays ($D^{*+}$ sample), and pions from kaon decays ($D^{*+}$ sample). We reweight the $\mathcal{R}(e/\mu)$ distribution for both the signal and background according to the weights obtained from the corresponding sample and then apply a new PDF for toy MC sample generation. The obtained systematic uncertainty is $0.13$.

To identify muons, we use PID with all four pairs of particle hypotheses (muon against electron, pion, kaon, or proton) in the BDT. To simplify, we evaluate the systematics of PID for all of them separately. Although they are correlated, the separate analysis only increases the systematic uncertainty.

For all kink mother particle types except for the muon, we have a clean sample providing the corresponding PID distribution in the data (electrons from the $\gamma$-conversion sample, kaons and pions from the $D^{*+}$ sample). Muons do not have a suitable sample; therefore, we treat them as pions instead. This replacement is justified since we use only $dE/dx$ losses, and they are almost the same for the muon and pion mass hypotheses. 

We reweight the $\mathcal{R}(\mu/x)$ distribution ($x=e$, $\pi$, $K$, or $p$) for both the signal and background and then apply a new PDF for toy MC sample generation. The obtained uncertainties are $0.13$ for $\mathcal{R}(\mu/e)$, $0.09$ for $\mathcal{R}(\mu/\pi)$, $0.10$ for $\mathcal{R}(\mu/K)$, and $0.06$ for $\mathcal{R}(\mu/p)$.

The combined PID in BDT systematic uncertainty is $0.24$.

\subsection{Signal PDF systematics}
Here, we study all sources of systematic uncertainties related to the signal PDF $\mathcal{P}_\text{sig}$. These include signal reconstruction efficiency depending on the muon laboratory momentum $p_\mu$ and the electron emission angle in the muon rest frame $\theta_{e\mu}$, electron momentum resolution in the muon rest frame, and also the systematics of the signal MC sample generation method.

The systematic uncertainty due to the discrepancy in reconstruction efficiency between the data and the MC simulation consists of two different contributions: one is related to the trigger efficiency of the selected topology, and the other is due to the kink reconstruction efficiency. The trigger efficiency uncertainty results in a small discrepancy between primary muon momentum distributions in the MC and data samples. We obtain weights for the estimation from the sample of $\tmnn$ decays without $\menn$ kink. This systematic uncertainty is $0.08$.

The reconstruction efficiency also strongly depends on the electron emission angle $\theta_{e\mu}$. We control this effect using the largest selected background sample of $\kmunu$ decays (from $D^{*+}$ decays). Since kaons are pseudoscalars, their decay angular distribution is uniform. Therefore, after reconstruction, the $\cos{\theta_{\mu K}}$ distribution represents the kink reconstruction efficiency. Applying weights from the kaon decay to our signal, we estimate the systematic uncertainty to be $0.09$.

To take into account the systematics of the electron momentum resolution in the muon rest frame, we also exploit the $\kmunu$ decay sample. We use kaons selected from the $\tau$ sample since here we observe a discrepancy in $p_{\mu K}$ resolution slightly larger compared to kaon kinks from the $D^{*+}$ sample. We estimate systematic uncertainty to be $0.03$.

The systematic uncertainty induced by the muon lifetime reduction for the signal MC sample generation is $0.07$. It is estimated by comparing the signal MC sample to the ten times smaller MC sample generated with the default muon lifetime.

The combined signal PDF systematic uncertainty is $0.14$.

\subsection{Fit procedure systematics}\label{sec:7.4}
To estimate the systematic uncertainty of the fit procedure, we compare the fit results of the Michel parameter $\xip$ for two different $\mathcal{P}_\text{sig}$. The first one is the default signal PDF in the form of a histogram. The second one is obtained from the MC $(y,\,\cos{\theta_e})$ distribution by smoothing a $16\times 16$-bin 2D histogram with a parametric function. We estimate the systematic uncertainty of this source to be $0.25$.

In addition, we check the difference in the result by varying the bin width of the default $\mathcal{P}_\text{sig}$. It is impossible to vary the bin size much since with too fine binning, a few empty bins appear, leading to a bias of the fit, while with too rough binning, the sensitivity suffers due to the fitting function sharpness in some regions. Thus, $8\times8$ and $12\times12$ net is used to check the variation. The obtained difference is small and does not exceed $0.13$.

In conclusion, we consider $0.25$ as a systematic uncertainty of the fit procedure. 

\subsection{Systematics summary}
Finally, the combined overall systematic uncertainty of the Michel parameter $\xip$ measurement is estimated to be $\sigma_{\xip}=0.42$. In Table~\ref{tab:7.1}, we summarize the results of the systematic uncertainty estimation for all sources.

\begin{table}[htb]
\caption{ Sources of the systematic uncertainties of the Michel parameter $\xip$ measurement (absolute values).}
\label{tab:7.1}
\begin{tabular}
 {@{\hspace{0.5cm}}l@{\hspace{0.5cm}}  @{\hspace{0.5cm}}c@{\hspace{0.5cm}}}
\hline \hline
Source & Uncertainty \\
\hline
\multicolumn{2}{c}{Background}\\
Purity ($p$) & $0.10$ \\
$\pi^-\to\mu^-\bar{\nu}_\mu$ MC & $0.05$ \\
$K^-\to3$ body MC  & $0.06$ \\
$K^-\to\mu^-\bar{\nu}_\mu$ MC  & $0.05$ \\
$e$-scattering MC & $0.10$ \\
hadron scattering MC & $0.11$ \\
\multicolumn{2}{c}{PID in BDT}\\
$\mathcal{R}(e/\mu)$ & 0.13 \\
$\mathcal{R}(\mu/e)$ & 0.13 \\
$\mathcal{R}(\mu/\pi)$ & 0.09 \\
$\mathcal{R}(\mu/K)$ & 0.10 \\
$\mathcal{R}(\mu/p)$ & 0.06 \\
\multicolumn{2}{c}{Signal PDF}\\
$p_\mu$ efficiency & 0.08 \\
$\cos{\theta_{e\mu}}$ efficiency & 0.09 \\
$p_{e\mu}$ resolution & 0.03 \\
Signal MC generation & 0.07 \\
\multicolumn{2}{c}{Fit procedure}\\
Fit function & 0.25 \\
\hline
Total & $0.42$ \\
\hline \hline
\end{tabular}
\end{table}

Systematic uncertainty is significantly smaller than the statistical one in this analysis, demonstrating the potential of the applied method. The method allows for a significant improvement in accuracy in the near future in already working experiments or those being under development. Thus, the task of control of systematic uncertainty with increasing statistics is worth considering. 

A qualitative consideration that the statistical uncertainty will dominate the systematic one in similar analyses in the near future experiments is discussed in detail in Ref.~\cite{Bodrov:2022mbd}. This analysis confirms in practice the validity of that conclusion: most of the systematic sources are controlled with large independent samples, and no limiting factors for further improvements in accuracy have yet been observed.

\section{Result}
We measure the Michel parameter $\xip$ to be
\begin{eqnarray}
\xip=\res,
\end{eqnarray}
where the first uncertainty is statistical and the second one is systematic. This result is consistent with the Standard Model prediction of $\xip_\text{SM}=1$. The combined uncertainty, $\sigma_{\xip}=1.03$, is more than two times smaller compared to the previous Belle result $\xip=-2.2\pm2.4$ calculated from $\xi\kappa$ obtained in the study of the $\tmnng$ decay~\cite{Shimizu:2017dpq}.

Based on the gained experience, it is possible to improve the result in the near future in the Belle~II experiment~\cite{Kou:2018nap}, taking into account the upgraded detector with an enlarged CDC and the implementation of improved tracking algorithms. In particular, the kink reconstruction algorithm implementation will provide a better momentum resolution, which is important for both background suppression and sensitivity increase (smeared by the resolution otherwise).

\section{Conclusion}
In summary, we report the first direct measurement of the Michel parameter $\xip$ in the $\tmnn$ decay with the full Belle data sample using the $\menn$ decay-in-flight in the Belle drift chamber. The obtained value of $\xip=\res$, where the first uncertainty is statistical and the second one is systematic, is in agreement with the Standard Model prediction $\xip_\text{SM}=1$. 

\section*{ACKNOWLEDGMENTS}
This work, based on data collected using the Belle detector, which was
operated until June 2010, was supported by 
the Ministry of Education, Culture, Sports, Science, and
Technology (MEXT) of Japan, the Japan Society for the 
Promotion of Science (JSPS), and the Tau-Lepton Physics 
Research Center of Nagoya University; 
the Australian Research Council including grants
DP210101900, 
DP210102831, 
DE220100462, 
LE210100098, 
LE230100085; 
Austrian Federal Ministry of Education, Science and Research (FWF) and
FWF Austrian Science Fund No.~P~31361-N36;
the National Natural Science Foundation of China under Contracts
No.~11675166,  
No.~11705209;  
No.~11975076;  
No.~12135005;  
No.~12175041;  
No.~12161141008; 
Key Research Program of Frontier Sciences, Chinese Academy of Sciences (CAS), Grant No.~QYZDJ-SSW-SLH011; 
Project ZR2022JQ02 supported by Shandong Provincial Natural Science Foundation;
the Ministry of Education, Youth and Sports of the Czech
Republic under Contract No.~LTT17020;
the Czech Science Foundation Grant No. 22-18469S;
Horizon 2020 ERC Advanced Grant No.~884719 and ERC Starting Grant No.~947006 ``InterLeptons'' (European Union);
the Carl Zeiss Foundation, the Deutsche Forschungsgemeinschaft, the
Excellence Cluster Universe, and the VolkswagenStiftung;
the Department of Atomic Energy (Project Identification No. RTI 4002) and the Department of Science and Technology of India; 
the Istituto Nazionale di Fisica Nucleare of Italy; 
National Research Foundation (NRF) of Korea Grant
Nos.~2016R1\-D1A1B\-02012900, 2018R1\-A2B\-3003643,
2018R1\-A6A1A\-06024970, RS\-2022\-00197659,
2019R1\-I1A3A\-01058933, 2021R1\-A6A1A\-03043957,
2021R1\-F1A\-1060423, 2021R1\-F1A\-1064008, 2022R1\-A2C\-1003993;
Radiation Science Research Institute, Foreign Large-size Research Facility Application Supporting project, the Global Science Experimental Data Hub Center of the Korea Institute of Science and Technology Information and KREONET/GLORIAD;
the Polish Ministry of Science and Higher Education and 
the National Science Center;
the Ministry of Science and Higher Education of the Russian Federation, Agreement 14.W03.31.0026, 
and the HSE University Basic Research Program, Moscow; 
University of Tabuk research grants
S-1440-0321, S-0256-1438, and S-0280-1439 (Saudi Arabia);
the Slovenian Research Agency Grant Nos. J1-9124 and P1-0135;
Ikerbasque, Basque Foundation for Science, Spain;
the Swiss National Science Foundation; 
the Ministry of Education and the Ministry of Science and Technology of Taiwan;
and the United States Department of Energy and the National Science Foundation.
These acknowledgements are not to be interpreted as an endorsement of any
statement made by any of our institutes, funding agencies, governments, or
their representatives.
We thank the KEKB group for the excellent operation of the
accelerator; the KEK cryogenics group for the efficient
operation of the solenoid; and the KEK computer group and the Pacific Northwest National
Laboratory (PNNL) Environmental Molecular Sciences Laboratory (EMSL)
computing group for strong computing support; and the National
Institute of Informatics, and Science Information NETwork 6 (SINET6) for
valuable network support.

\appendix

\section{Kink events selection in the decay chain \boldmath{$D^{*+}\to D^0(\to K^-\pi^+)\pi^+$}}
\label{app:1}

We reconstruct $D^{*+}$ candidates in the decay chain $D^{*+}\to D^0(\to K^-\pi^+)\pi^+$. The following selection criteria on the $K^-\pi^+$ and $K^-\pi^+\pi^+$ invariant masses are used: $1.82\,\gevcc<M(K^-\pi^+)<1.9\,\gevcc$ and $|M(K^-\pi^+\pi^+)-M(K^-\pi^+) + M^\text{PDG}(D^0) - M^\text{PDG}(D^{*+})|<3\,\text{MeV}/c^2$, providing a large sample of $D^{*+}$ candidates. The momentum of the $D^{*+}$ candidates in the c.m. frame is limited at $p_{D^{*+}}>2.3\,\gevc$ since our MC simulation of $\ee\to\Upsilon(4S)\to B\bar{B}$ does not reproduce $K^-\pi^+\pi^+$ invariant mass distribution well for both the $D^{*+}$ signal and combinatorial background. For larger momentum, $D^{*+}$ are produced in the continuum, and our MC simulation of $\ee\to\qq$ describes the combinatorial background well. However, there is a discrepancy between the data and MC samples in the $D^{*+}$ peak since the effects of the $c$-quark fragmentation were not properly accounted for in the MC simulation. The fragmentation is based mainly on the momentum spectrum; therefore, we reweight the MC sample with a real $D^{*+}\to D^0\pi^+$ decay in bins of its momentum. The following procedure is used: the $M(K^-\pi^+\pi^+)$ distribution in data is fitted in bins of $p_{D^{*+}}$, and the number of $D^{*+}$ mesons is obtained. After that, we determine the weight for the MC event with real $D^{*+}$ as $w(p_{D^{*+}})=N^\text{data}_{D^{*+}}(p_{D^{*+}})/N^\text{MC}_{D^{*+}}(p_{D^{*+}})$. We perform this procedure with $D^{*+}$ candidates reconstructed before any kink selection.

For further event selection, we require one of the $D^0$ daughter tracks to pass the kink selection criteria described in Sec.~\ref{sec:4.2}. The second track is identified using the information from the CDC, TOF, and ACC combined to form likelihood $\mathcal{L}_i$ ($i=\pi$ or $K$). To select the pion (kaon) kink, we require $\mathcal{R}(K/\pi)=\mathcal{L}_{K}/(\mathcal{L}_{K}+\mathcal{L}_\pi)>0.6$ [$\mathcal{R}(\pi/K)>0.6$] for $K^-$ ($\pi^+$) from the $D^0$ meson.

\begin{figure}[!hbp]
\includegraphics[width=1\linewidth]{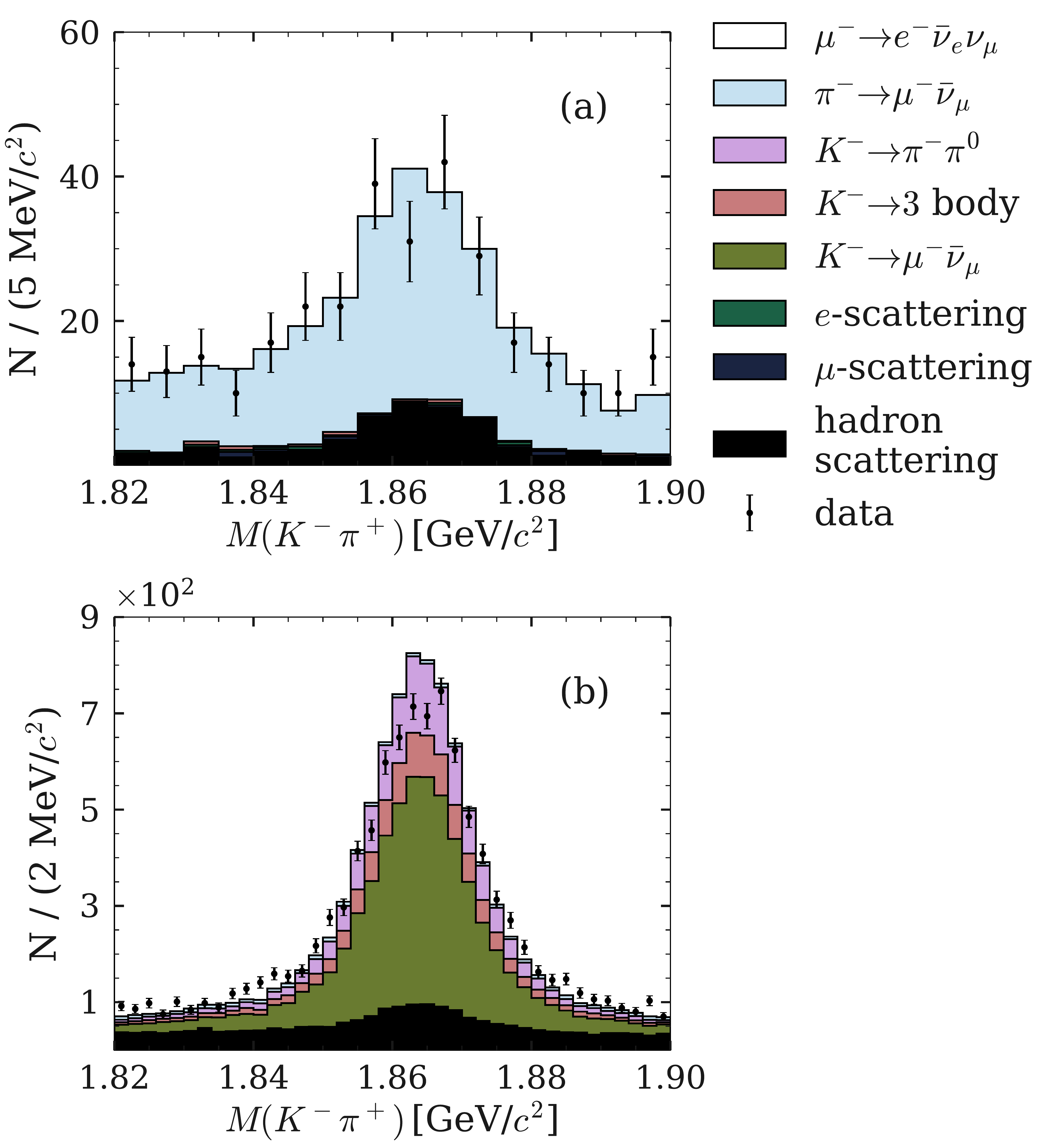}
\caption{$K^-\pi^+$ invariant mass for (a) $\pi^+$ kink candidates and (b) $K^-$ kink candidates.}
\label{fig:app1}
\end{figure}

To illustrate the result of the selection, we plot the $K^-\pi^+$ invariant mass for pion and kaon kinks in Fig.~\ref{fig:app1}(a) and~(b), respectively. As can be seen, each sample consists of the corresponding kinks, as well as hadron scattering events. 

\section{Kink events selection in the \boldmath{$\gamma$}-conversion process}
\label{app:2}

We select $\gamma$-conversion events from 1--1 and 1--3 topology $\tat$ pairs sample prepared according to the preselection criteria described in Sec.~\ref{sec:4.1}. Although this preselection limits available statistics, the kink reconstruction efficiency here is similar to one in the main analysis. 

The conversion is reconstructed on the one-track side from two oppositely charged tracks. To suppress background from other $V$-shaped processes like $K^0_S$ decay, the invariant mass of $\ee$ pair $m_{\ee}$ is required to be less than $40\,\text{MeV}/c^2$. Since $\gamma$-conversion occurs on the detector material, the radius of the conversion vertex in the $r\phi$-plane has to be larger than $2\,\text{cm}$. To suppress a random combination of the tracks, the distance between two tracks in projection onto the $z$-axis is required to be less than $5\,\text{cm}$. The daughter electron is reconstructed as a kink with the selection criteria described in Sec.~\ref{sec:4.2}. Finally, using the identification of the daughter positron $\mathcal{R}(e/x)>0.8$, we obtain a clean sample of identified electron scattering events.

\begin{figure}[!hbp]
\includegraphics[width=1\linewidth]{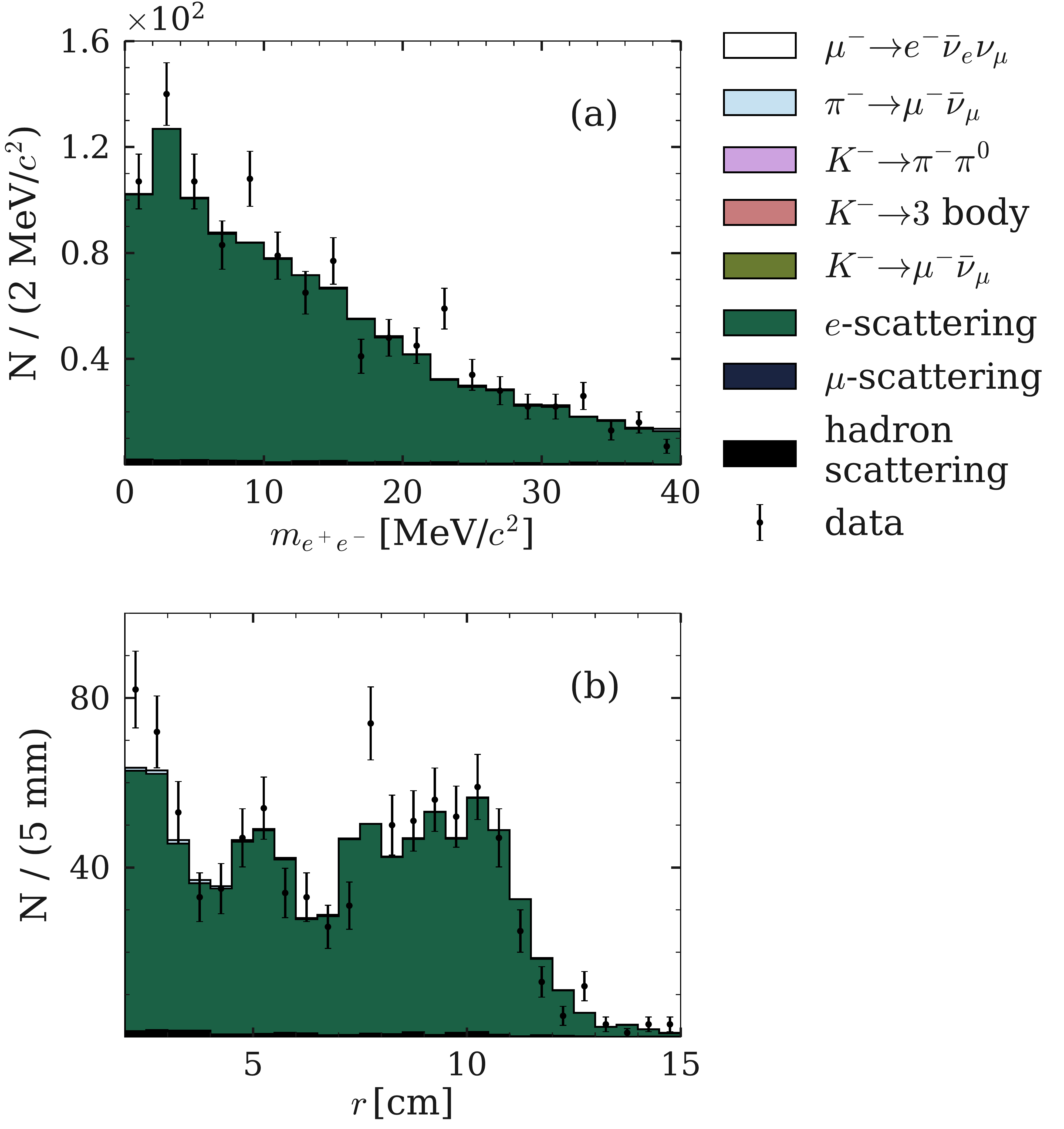}
\caption{(a) invariant mass $m_{\ee}$ and (b) radius of the conversion vertex for the selected $\gamma$-conversion events, where one of the electron is reconstructed as an electron scattering kink.}
\label{fig:app2}
\end{figure}

To illustrate the result of the described procedure, we plot the $\ee$-pair invariant mass in Fig.~\ref{fig:app2}(a) and the radius of the conversion vertex in the $r\phi$-plane in Fig.~\ref{fig:app2}(b). The localization of the $m_{\ee}$ in the zero region is as expected. In the distribution of the $\gamma$-conversion vertex, the SVD structure is clearly observed. The selected sample consists of pure electron scattering events.

\clearpage

\bibliography{bibl}

\end{document}